\documentclass[twocolumn,showpacs,preprintnumbers,amsmath,amssymb,floatfix,prb]{revtex4}

\usepackage{graphicx}
\usepackage{dcolumn}
\usepackage{bm}
\usepackage{subfigure}

\begin{document}

\title{Epitaxial strain effects in the spinel ferrites CoFe$_2$O$_4$ and NiFe$_2$O$_4$ from first principles}

\author{Daniel Fritsch}
\email{fritschd@tcd.ie}
\author{Claude Ederer}
\affiliation{School of Physics, Trinity College Dublin, Dublin 2, Ireland}

\date{\today}

\begin{abstract}
The inverse spinels CoFe$_{2}$O$_{4}$ and NiFe$_{2}$O$_{4}$, which
have been of particular interest over the past few years as building
blocks of artificial multiferroic heterostructures and as possible
spin-filter materials, are investigated by means of density functional
theory calculations. We address the effect of epitaxial strain on the
magneto-crystalline anisotropy and show that, in agreement with
experimental observations, tensile strain favors perpendicular
anisotropy, whereas compressive strain favors in-plane orientation of
the magnetization. Our calculated magnetostriction constants
$\lambda_{100}$ of about $-220$~ppm for CoFe$_2$O$_4$ and $-45$~ppm for
NiFe$_2$O$_4$ agree well with available experimental data. We analyze the
effect of different cation arrangements used to represent the inverse
spinel structure and show that both LSDA+$U$ and GGA+$U$ allow for a
good quantitative description of these materials. Our results open the
way for further computational investigations of spinel ferrites.
\end{abstract}



\maketitle

\section{Introduction}
\label{Introduction}

Spinel ferrites CoFe$_2$O$_4$ and NiFe$_2$O$_4$ are insulating
magnetic oxides with high magnetic ordering temperatures and large
saturation magnetizations.\cite{Brabers1995189} This rare combination of
properties makes them very attractive for a wide range of
applications. Recently, particular attention has been focused on the
possible use of spinel ferrites as magnetic components in artificial
multiferroic
heterostructures\cite{Zheng_Science303_661,Zavaliche_et_al:2005,Dix_MatSciEngB144_127,Muralidharan_JAP103_07E301}
or as spin-filtering tunnel barriers for spintronics
devices.\cite{Lueders_JAP99_08K301,Chapline/Wang:2006,Ramos_PRB78_180402}

For these applications, the corresponding materials have to be
prepared either in the form of thin films, grown on different
substrates, or as components of more complex epitaxial
heterostructures.\cite{Suzuki:2001,Lueders_PRB71_134419,Zheng_Science303_661,Zhou_PRB80_094409}
Due to the mismatch in lattice constants and thermal expansion coefficients between the thin film material and the
substrate, significant amounts of strain can be incorporated in such
epitaxial thin film structures, depending on the specific growth
conditions and substrate materials. This epitaxial strain
can then lead to drastic changes in the properties of the thin film
material. Indeed, a reoriention of the magnetic easy axis under
different conditions has been reported for
CoFe$_2$O$_4$,\cite{Huang_APL89_265206,Lisfi_PRB76_054405,Gao_JPhysD42_175006}
and a strong enhancement of magnetization and conductivity has been
observed in NiFe$_2$O$_4$ thin
films.\cite{Lueders_PRB71_134419,Lueders_AdvMat18_1733,Rigato_MatSciEngB144_43}

In order to efficiently optimize the properties of thin film
materials, it is important to clarify whether the observed deviations
from bulk behavior are indeed due to the epitaxial strain or whether
they are induced by other factors, such as for example defects,
off-stoichiometry, or genuine interface effects. First principles
calculations based on density functional theory
(DFT),\cite{Hohenberg/Kohn:1964,Kohn/Sham:1965,Jones/Gunnarsson:1989} can provide
valuable insights in this respect by allowing to address each of these effects separately.

Here we present results of DFT calculations for the structural and
magnetic properties of epitaxially strained CoFe$_2$O$_4$ and
NiFe$_2$O$_4$, with special emphasis on strain-induced changes in the
magneto-crystalline anisotropy energy (MAE). Our results are
representative for (001)-oriented thin films of CoFe$_2$O$_4$ and
NiFe$_2$O$_4$, grown on different lattice-mismatched substrates. Our
results provide important reference data for the interpretation of
experimental observations in spinel ferrite thin films and in
heterostructures consisting of combinations of spinel ferrites with
other materials, such as perovskite structure oxides.

We find a large and strongly strain-dependent MAE for CoFe$_2$O$_4$,
and a smaller but also strongly strain-dependent MAE for
NiFe$_2$O$_4$. We discuss the influence of different cation arrangements within the
inverse spinel structure and analyze the difference in the structural and magnetic properties due to different
exchange-correlation functionals used in the calculations. From our calculations we obtain the magnetostriction
constants $\lambda_{100}$ for both CoFe$_2$O$_4$ and
NiFe$_2$O$_4$, which agree well with available experimental data.

This paper is organized as follows. Sec.~\ref{TheorySpinelStructure}
gives a brief overview over the properties of CoFe$_2$O$_4$ and NiFe$_2$O$_4$ that are
important for the present work and also summarizes results of previous
DFT calculations. The basic equations governing the magnetoelastic
properties of cubic crystals are presented in
Sec.~\ref{TheoryMagnetoelasticEnergyAndMAE}, followed by a detailed
description of the structural relaxations performed in this work in
Sec.~\ref{TheoryStructuralRelaxations}, and a summary of further
computational details in Sec.~\ref{TheoryComputationalDetails}. The
results of the bulk structural properties will be presented in
Sec.~\ref{ResultsBulkStructureAndDOS}, whereas the effect of strain on the structural properties is analyzed in Sec.~\ref{ResultsStrainAndElasticProperties}. The effect of strain on the MAE is discussed in Sec.~\ref{ResultsMAEAndMagnetoelasticProperties}. Finally, in Sec.~\ref{SummaryAndOutlook} a summary of our main conclusions is given.

\section{Background and computational details}
\label{Theory}

\subsection{Spinel structure and previous work on ferrites}
\label{TheorySpinelStructure}

\begin{figure}[b]
\subfigure[][]{
\label{SpinelStructureInverseRandom}
\includegraphics[width=0.5\textwidth,clip]{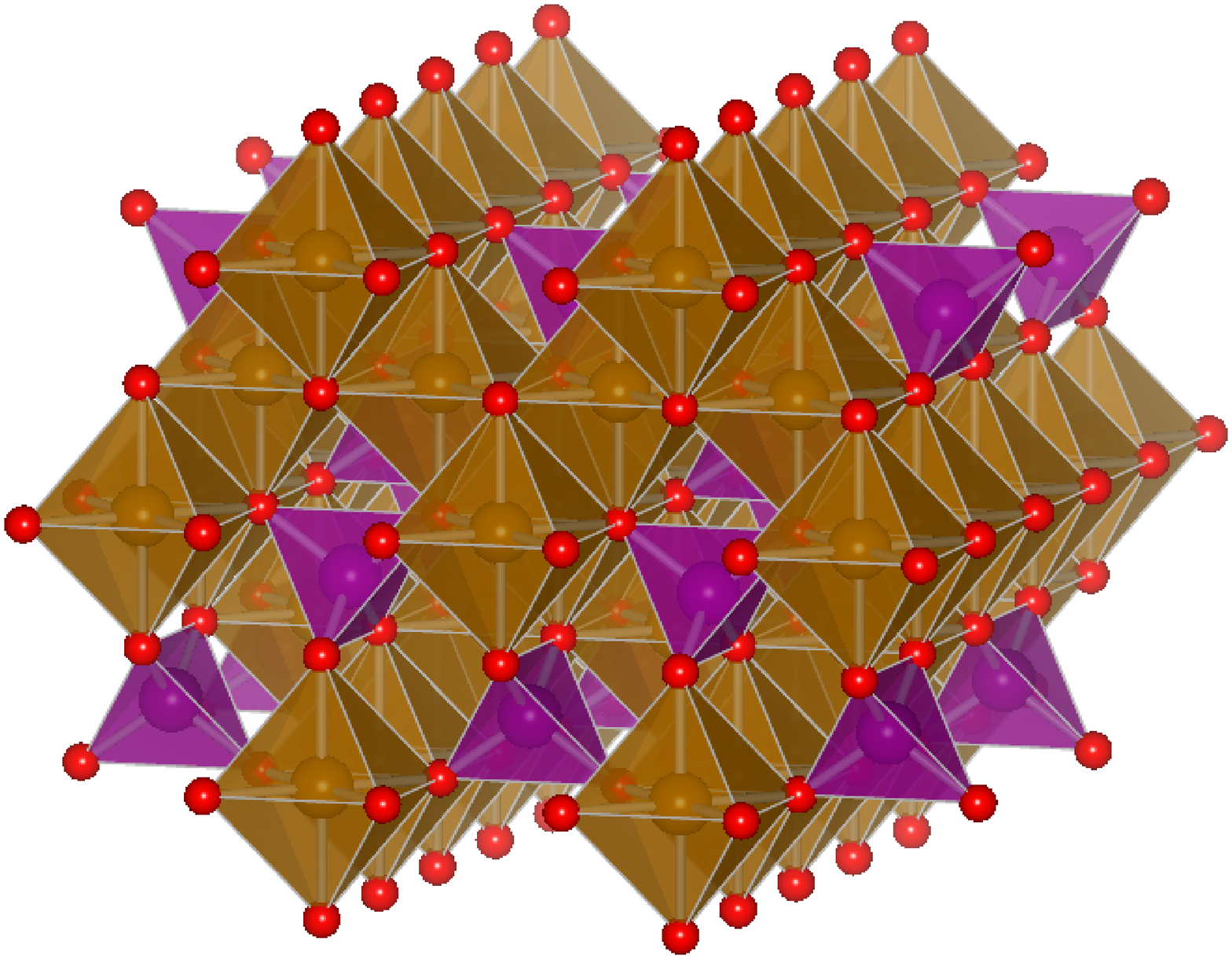}
}
\subfigure[][]{
\label{SpinelStructureInverse}
\includegraphics[width=0.5\textwidth,clip]{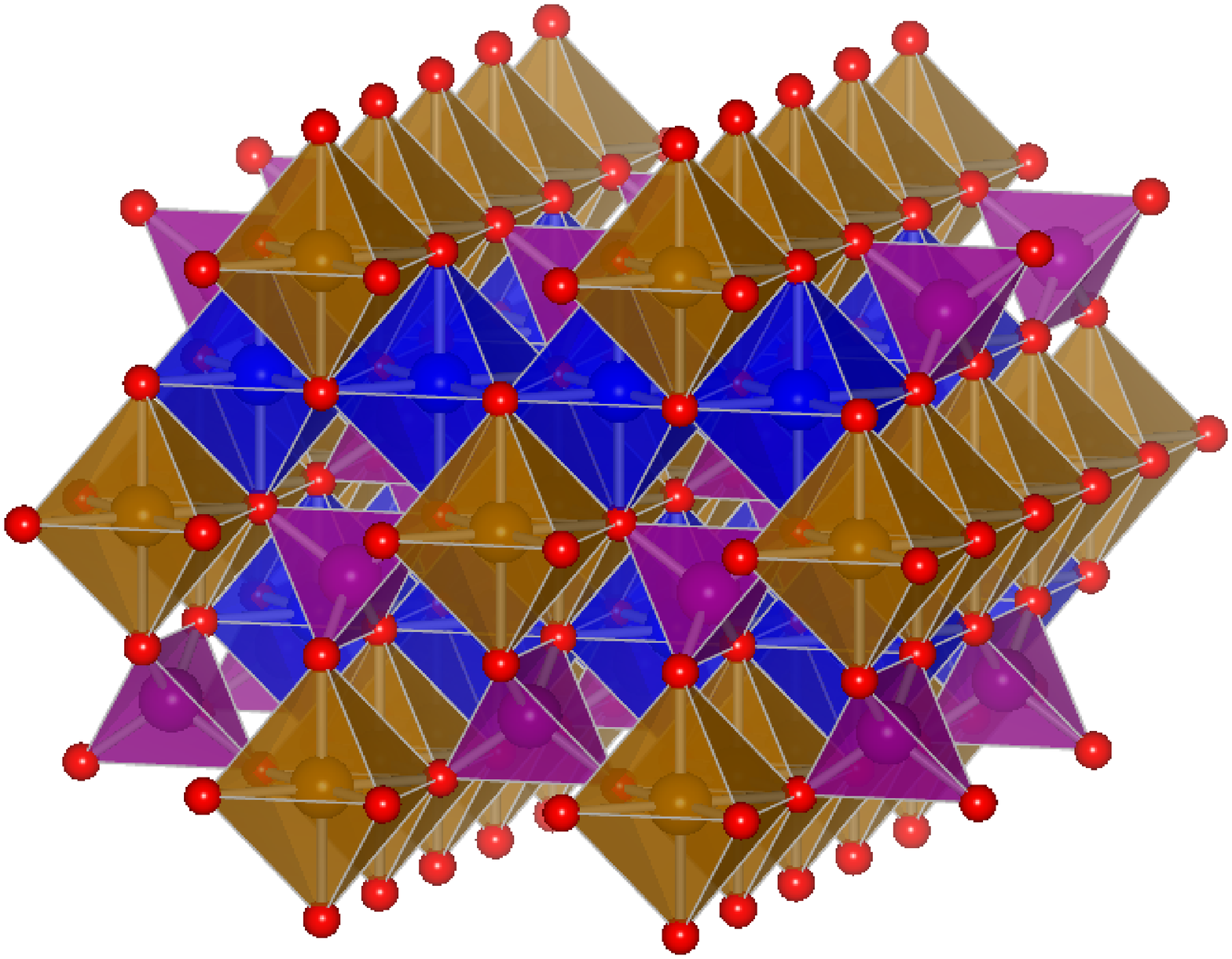}
}
\caption{\label{SpinelStructure}(Color online) Inverse spinel
  structure of spinel ferrites. Oxygen cations are depicted as small
  red spheres, and the coordination polyhedra surrounding cation sites
  are shaded. \subref{SpinelStructureInverseRandom} Random cation
  distribution on the octahedrally coordinated $B$ sites, i.e. all $B$
  sites remain equivalent. \subref{SpinelStructureInverse} $Imma$
  cation arrangement used throughout this work. Inequivalent $B$ sites
  are represented by different shadings of the corresponding
  octahedra.}
\end{figure}

The spinel structure (space group $Fd\bar{3}m$, general formula
$AB_{2}X_{4}$) contains two inequivalent cation sites, the
tetrahedrally-coordinated $A$ site ($T_d$ symmetry, Wyckoff position
8a), and the octahedrally-coordinated $B$ site ($O_h$ symmetry,
Wyckoff position 16d). In the \emph{normal spinel} structure, all $A$
sites are occupied by one cation species (divalent cation), whereas
all $B$ sites are occupied by the other cation species (trivalent
cation).  On the other hand, in the \emph{inverse spinel} structure
the trivalent cations occupy all $A$ sites as well as 50~\% of the $B$
sites, whereas the remaining 50~\% of the $B$ sites are occupied by
the divalent cations. If the distribution of divalent and trivalent
cations on the $B$ sites is completely random, all $B$ sites remain
crystallographically equivalent and the overall cubic $Fd\bar{3}m$
symmetry is preserved (see Fig.~\ref{SpinelStructureInverseRandom}).

Both CoFe$_2$O$_4$ and NiFe$_2$O$_4$ crystallize in the inverse spinel
structure, even though for CoFe$_2$O$_4$ the inversion is typically
not fully complete, i.e. there is a nonzero Co$^{2+}$ occupation on
the $A$ site. Thereby, the exact degree of inversion depends strongly
on the preparation conditions.\cite{Brabers1995189} To the best of our
knowledge, no deviations from cubic symmetry have been reported for
either system, i.e. the Co$^{2+}$/Ni$^{2+}$ cations are believed to be
randomly distributed on the $B$ sites.\cite{NiFe2O4}

According to the formal $d^5$ and $d^7$ electron configurations
corresponding to Fe$^{3+}$ and Co$^{2+}$, respectively, these ions can
in principle exhibit both high-spin and low-spin states, but only the
high-spin states are experimentally observed in both spinel ferrites.
For the $d^8$ electron configuration of Ni$^{2+}$ no
such distinction exists. The magnetic moments of the $A$ site cations are
oriented antiparallel to the
magnetic moments of the $B$ site cations: the so-called N{\'e}el-type
ferrimagnetic arrangement.\cite{Neel} Thus, the magnetic moments
of the Fe$^{3+}$ cations on the $A$ and $B$ sites cancel each other
exactly, and the net magnetization is mainly due to the divalent
$B$ site cations, i.e. either Co$^{2+}$ or Ni$^{2+}$. This results in a
magnetic moment per formula unit close to the formal values of 3
$\mu_B$ and 2 $\mu_B$ in CoFe$_2$O$_4$ and NiFe$_2$O$_4$,
respectively. Deviations from these values can be either due to
orbital contributions to the magnetic moments, or due to incomplete
inversion and off-stoichiometric cation distribution.

Both CoFe$_2$O$_4$ and NiFe$_2$O$_4$ are small gap insulators, but information on the
experimental gap size is very limited. Waldron used infrared spectra
to obtain threshold values of 0.11~eV and 0.33~eV for the electronic
transitions in CoFe$_2$O$_4$ and NiFe$_2$O$_4$, respectively,
\cite{Waldron_PhysRev99_1727} whereas Jonker estimated the energy gap
in CoFe$_2$O$_4$ to be 0.55~eV, based on resistivity measurements
along with other methods.\cite{Jonker_JPhysChemSol9_165}

Theoretical calculations of the electronic structure of spinel
ferrites so far have been focused mostly on magnetite
(Fe$_3$O$_4$). This material can be viewed as parent compound for the
spinel ferrites, including CoFe$_2$O$_4$ and NiFe$_2$O$_4$, which are
obtained by substituting the Fe$^{2+}$ cation in magnetite by a
different divalent 3$d$ transition metal cation. In an early work,
P\'{e}nicaud \textit{et al.} performed DFT calculations within the
local spin-density approximation (LSDA) for magnetite and the
respective Co-, Ni-, Mn-, and Zn-substituted
ferrites.\cite{Penicaud_JMagMagMat103_212} The use of LSDA leads to
half-metallic band-structures for all systems except NiFe$_2$O$_4$, in
contrast to the insulating character observed experimentally. (We note
that the case of magnetite is somewhat more involved than that of the
other spinel ferrites, since magnetite exhibits a metal-insulator
transition at $\sim$120~K.) It was later shown by Antonov \textit{et
  al.} that insulating solutions for Co-, Ni-, and Mn- substituted
Fe$_3$O$_4$ can be obtained within the LSDA+$U$
approach.\cite{Antonov_PRB67_024417} The same was found by Szotek
\textit{et al.} using a self-interaction-corrected LSDA
approach.\cite{Szotek_PRB74_174431} The latter study also addressed
the energetic difference between normal and inverse spinel structures
with different valence configurations. The electronic
structure of NiFe$_2$O$_4$ was also calculated within a hybrid functional
approach, where a large band-gap of 4~eV was obtained by using 40~\%
of Hartree-Fock exchange in the exchange-correlation energy
functional.\cite{Zuo_et_al:2006} Recently, Perron \textit{et al.}
investigated different magnetic arrangements for NiFe$_2$O$_4$ in both
normal and inverse spinel structures using both LSDA and the
generalized gradient approximation (GGA), and found the inverse spinel
structure with N{\'e}el-type ferrimagnetic order to be energetically
most favorable,\cite{Perron_JPCM19_346219} in agreement with the
experimental observations.

Calculations of the MAE in strained CoFe$_2$O$_4$ and NiFe$_2$O$_4$ have been reported
by Jeng and Guo.\cite{Jeng_JMagMagMat239_88,Jeng_JMagMagMat240_436}
However, due to the use of the LSDA, these calculations were based on
half-metallic band-structures for both materials. Furthermore, no
information on structural properties or the influence of
the specific cation arrangement used in the calculation were given.

\subsection{Magnetoelastic energy of a cubic crystal}
\label{TheoryMagnetoelasticEnergyAndMAE}

In this work we are concerned with the effect of epitaxial strain on
the structural and magnetic properties of CoFe$_2$O$_4$ and NiFe$_2$O$_4$, i.e. with
the elastic and magnetoelastic response of these systems. Here, we
therefore give a brief overview over the general magnetoelastic theory
for a cubic crystal, and present the most important equations that are
used in Sec.~\ref{ResultsAndDiscussion} to analyze the results of
our first principles calculations.

The magnetoelastic energy density $f=E/V$ of a cubic crystal can be
written as:~\cite{Kittel_RevModPhys21_541}
\begin{equation}
\label{EquationTotalEnergyDensity}
f = f_{K} + f_{\text{el}} + f_{\text{mel}}\,,
\end{equation}
where the three individual terms describe the cubic (unstrained)
magnetic anisotropy energy density $f_K$, the purely elastic energy
density $f_\text{el}$ and the coupled magnetoelastic contribution
$f_\text{mel}$, respectively. To lowest order in the strain tensor
$\varepsilon_{ij}$ and in the direction cosines $\alpha_i$ of the
magnetization vector, these terms have the following forms:
\begin{equation}
\label{EquationFaniso}
f_{K} = K(\alpha_{1}^{2}\alpha_{2}^{2} + \alpha_{2}^{2}\alpha_{3}^{2}
+ \alpha_{3}^{2}\alpha_{1}^{2})\,,
\end{equation}
\begin{equation}
\label{EquationFelast}
\begin{split}
f_{\text{el}} =& \frac{1}{2}C_{11}(\varepsilon_{xx}^{2} +
\varepsilon_{yy}^{2} + \varepsilon_{zz}^{2})\\ & +
\frac{1}{2}C_{44}(\varepsilon_{xy}^{2} + \varepsilon_{yz}^{2} +
\varepsilon_{zx}^{2})\\ & + C_{12}(\varepsilon_{yy}\varepsilon_{zz} +
\varepsilon_{xx}\varepsilon_{zz} + \varepsilon_{xx}\varepsilon_{yy})\,
,
\end{split}
\end{equation}
\begin{equation}
\label{EquationFmel}
\begin{split}
f_{\text{mel}} =& B_{1}(\alpha_{1}^{2}\varepsilon_{xx} +
\alpha_{2}^{2}\varepsilon_{yy} + \alpha_{3}^{2}\varepsilon_{zz})\\ & +
B_{2}(\alpha_{1}\alpha_{2}\varepsilon_{xy} +
\alpha_{2}\alpha_{3}\varepsilon_{yz} +
\alpha_{3}\alpha_{1}\varepsilon_{zx})\,.
\end{split}
\end{equation}
Here, $K$ denotes the lowest order cubic anisotropy constant,
$C_{11}$, $C_{12}$, and $C_{44}$ are the elastic moduli, and $B_{1}$
and $B_{2}$ are magnetoelastic coupling constants.

The bulk modulus $B$ is defined as:
\begin{equation}
B = V_{0} \left.\left(\frac{\partial^{2}E_{\text{tot}}}{\partial V^{2}}
\right)\right|_{(V=V_{0})}\,,
\end{equation}
where $E_{\text{tot}}$ is the total energy and $V_0$ is the
equilibrium bulk volume. Using
Eqs.~\eqref{EquationTotalEnergyDensity}-\eqref{EquationFmel} the
bulk modulus of a cubic crystal can be expressed in terms of the
elastic moduli $C_{11}$ and $C_{12}$:
\begin{equation}
\label{EquationBulkModulus}
B = \frac{1}{3}\left(C_{11}+2C_{12}\right)\,.
\end{equation}

In this work we investigate the effect of epitaxial strain that is
induced in thin film samples by the lattice mismatch to the
substrate. This situation can be described by a fixed in-plane strain
$\varepsilon_{xx}=\varepsilon_{yy}=(a-a_0)/a_0$, where $a$ is the
in-plane lattice constant of the thin film material and $a_0$ is the
corresponding lattice constant in the bulk. The resulting out-of-plane
strain $\varepsilon_{zz}=(c-a_0)/a_0$ can be obtained from
Eqs.~\eqref{EquationTotalEnergyDensity}-\eqref{EquationFmel} together
with the condition of vanishing stress for the out-of-plane lattice constant $c$,
i.e. $\frac{\partial f}{\partial
  \varepsilon_{zz}}=0$. In the demagnetized state $\varepsilon_{zz}$ is related to the applied
in-plane strain via the so-called two-dimensional Poisson ratio
$\nu_{2D}$:\cite{Harrison_QuantumWellsWiresAndDots}
\begin{equation}
\label{Equation2DPoissonRatio}
\nu_{2D} = - \frac{\varepsilon_{zz}}{\varepsilon_{xx}} =
2\frac{C_{12}}{C_{11}}\,.
\end{equation}
After calculating both $\nu_\text{2D}$ and the bulk modulus using DFT,
the elastic moduli $C_{11}$ and $C_{12}$ can thus be obtained from
Eqs.~\eqref{EquationBulkModulus} and \eqref{Equation2DPoissonRatio}.

In Sec.~\ref{ResultsMAEAndMagnetoelasticProperties} we monitor the
differences in total energy for different orientations of the
magnetization as a function of the in-plane constraint
$\varepsilon_{xx}$. Using expression~\eqref{Equation2DPoissonRatio}
for $\nu_{2D}$ and taking the energy for orientation of the
magnetization along the [001] direction as reference, i.e. $\Delta
f_{hkl} = f_{001} - f_{hkl}$, one obtains:
\begin{equation}
\label{EquationB1}
\begin{split}
\Delta f_{\langle100\rangle} & = - B_1 (\nu_{2D}+1)\, \epsilon_{xx}\,,\\
\Delta f_{\langle110\rangle} & = - B_1 (\nu_{2D}+1)\, \epsilon_{xx} - \frac{1}{4} K\,,\\
\Delta f_{\langle111\rangle} & = -\frac{2}{3} B_1 (\nu_{2D}+1)\, \epsilon_{xx} -
\frac{3}{4} K\,,\\
\Delta f_{\langle101\rangle} & = -\frac{1}{2} B_1 (\nu_{2D}+1)\, \epsilon_{xx} -
\frac{1}{4} K\,.
\end{split}
\end{equation}
Thus, the epitaxial strain-dependence of these energy differences is
governed by the magnetoelastic coupling constant $B_1$ and the
two-dimensional Poisson ratio $\nu_{2D}$.

Since the constant $B_1$ is not directly accessible by experiment, the
linear magnetoelastic response is typically
characterized by the magnetostriction constant $\lambda_{100}$, which
is related to $B_1$ and the elastic moduli $C_{11}$ and $C_{12}$:
\begin{equation}
\label{EquationLambda100}
\lambda_{100}=-\frac{2}{3}\frac{B_{1}}{C_{11}-C_{12}}\,.
\end{equation}
$\lambda_{100}$ characterizes the relative change in length (lattice
constant) along [100] when the material is magnetized along this
direction, compared to the unmagnetized state.

\subsection{Structural relaxations for the inverse spinel structure}
\label{TheoryStructuralRelaxations}

As described in Sec.~\ref{TheorySpinelStructure} the distribution of
divalent and trivalent cations on the octahedrally coordinated $B$
sites in the inverse spinel structure is assumed to be random for both
CoFe$_2$O$_4$ and NiFe$_2$O$_4$. On the other hand, the periodic
boundary conditions employed in our calculations always correspond to
a specific cation arrangement with perfect long-range order. Even
though a ``quasi-random'' distribution of divalent and trivalent
cations could in principle be achieved by using a very large unit
cell, the required computational effort would be prohibitively
large. For simplicity, we therefore restrict ourselves to the smallest
possible unit cell which contains two spinel formula units, i.e. four
$B$ sites. Distributing two Fe atoms on two of these sites, and
filling the other two sites with either Co or Ni, lowers the symmetry
from spacegroup $Fd\bar{3}m$ (\#227) to $Imma$ (\#74) independent of
which of the four B sites are occupied by Fe (see
Fig.~\ref{SpinelStructureInverse}). Different choices (``settings'')
simply lead to different orientations of the orthorhombic axes
relative to the Cartesian directions. It will become clear from the
results presented in Sec.~\ref{ResultsAndDiscussion} that the specific
cation arrangement used in our calculations does not critically affect
our conclusions.

Within the lower $Imma$ symmetry (and setting 1, see below), the tetrahedrally coordinated $A$
sites are located on Wyckoff position 4e $(0, \tfrac{1}{4}, z)$,
whereas the octahedrally coordinated $B$ sites split into Wyckoff
positions 4b $(0, 0, \tfrac{1}{2})$ and 4d $(\tfrac{1}{4},
\tfrac{1}{4}, \tfrac{3}{4})$. In addition, the oxygen positions split
into Wyckoff positions 8i $(x, \tfrac{1}{4}, z)$ and 8h $(0, y,
z)$.

\begin{figure}
\centering

\subfigure[][]{
\label{FigureInverseSpinelSetting1}
\begin{minipage}{0.225\textwidth}
\includegraphics[width=\textwidth,clip]{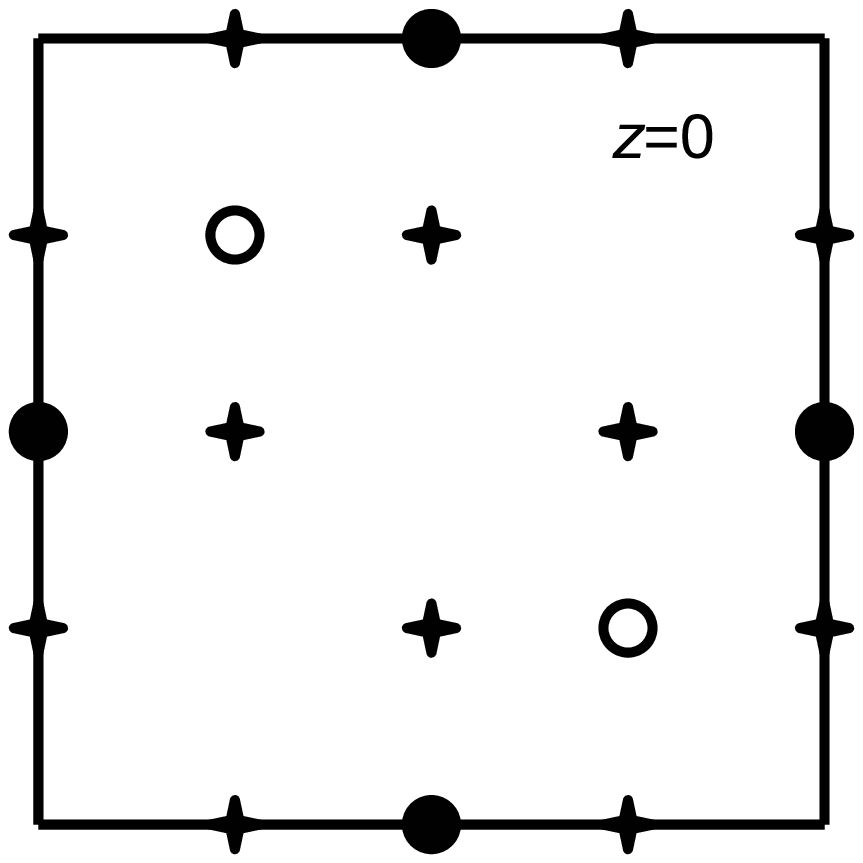}\\
\includegraphics[width=\textwidth,clip]{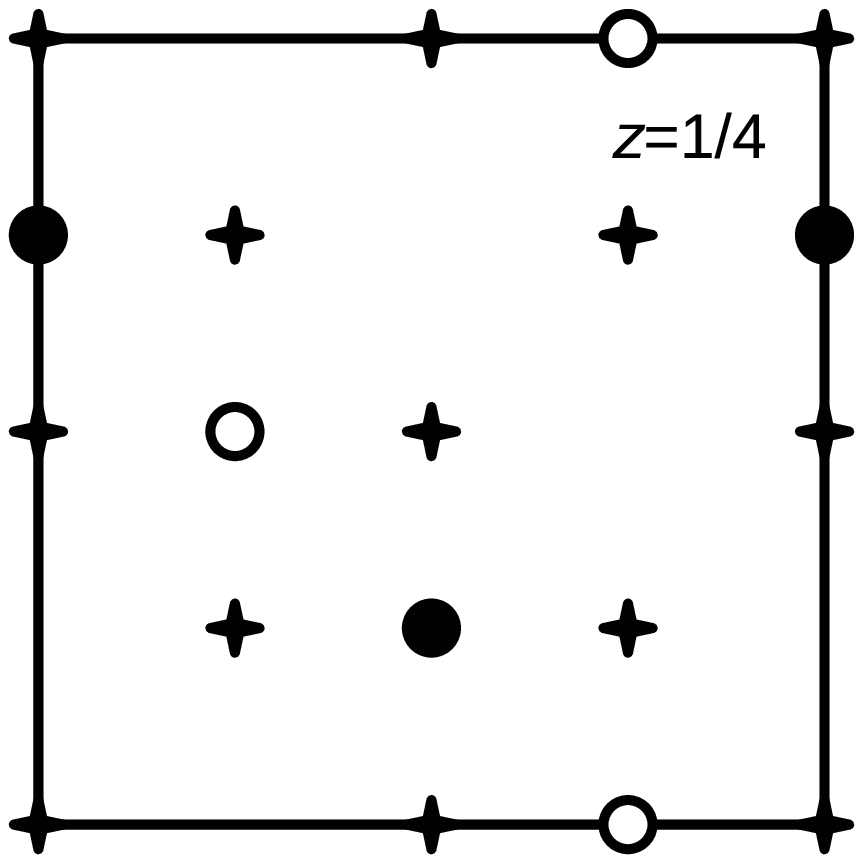}\\
\includegraphics[width=\textwidth,clip]{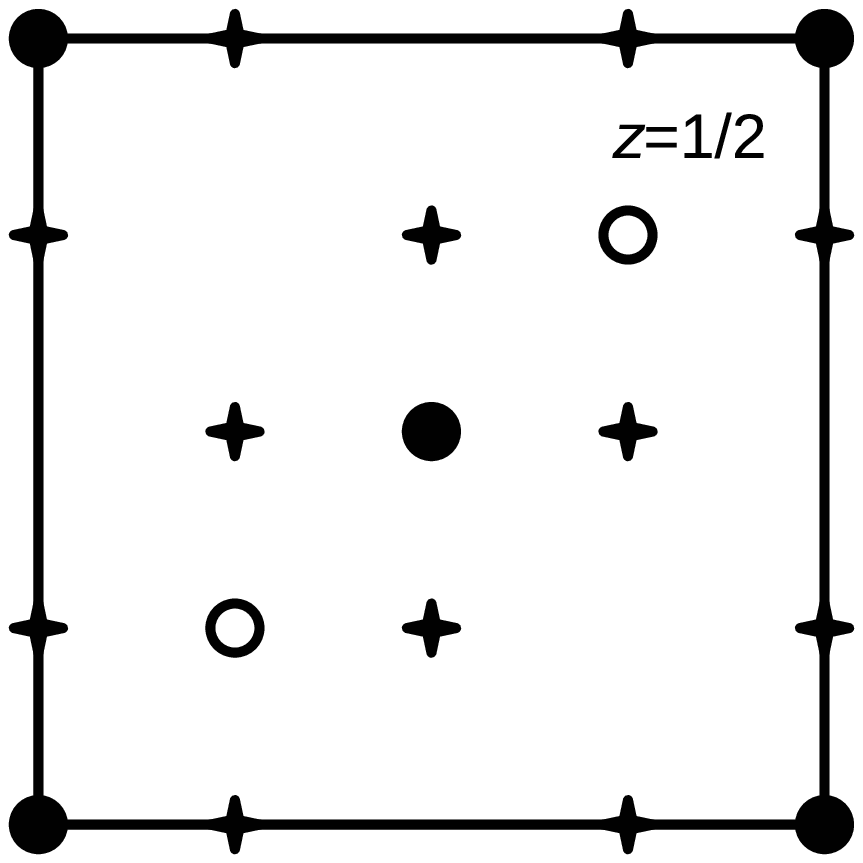}\\
\includegraphics[width=\textwidth,clip]{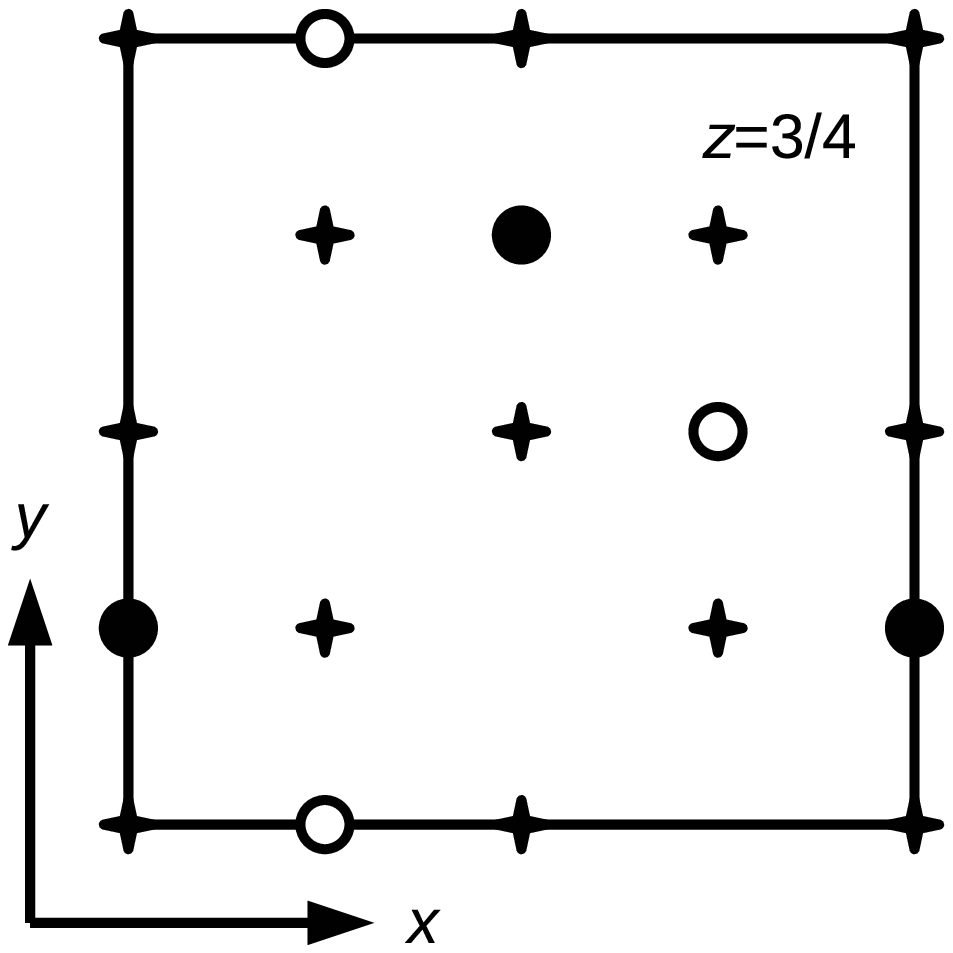}\\
\end{minipage}
}
\subfigure[][]{
\label{FigureInverseSpinelSetting3}
\begin{minipage}{0.225\textwidth}
\includegraphics[width=\textwidth,clip]{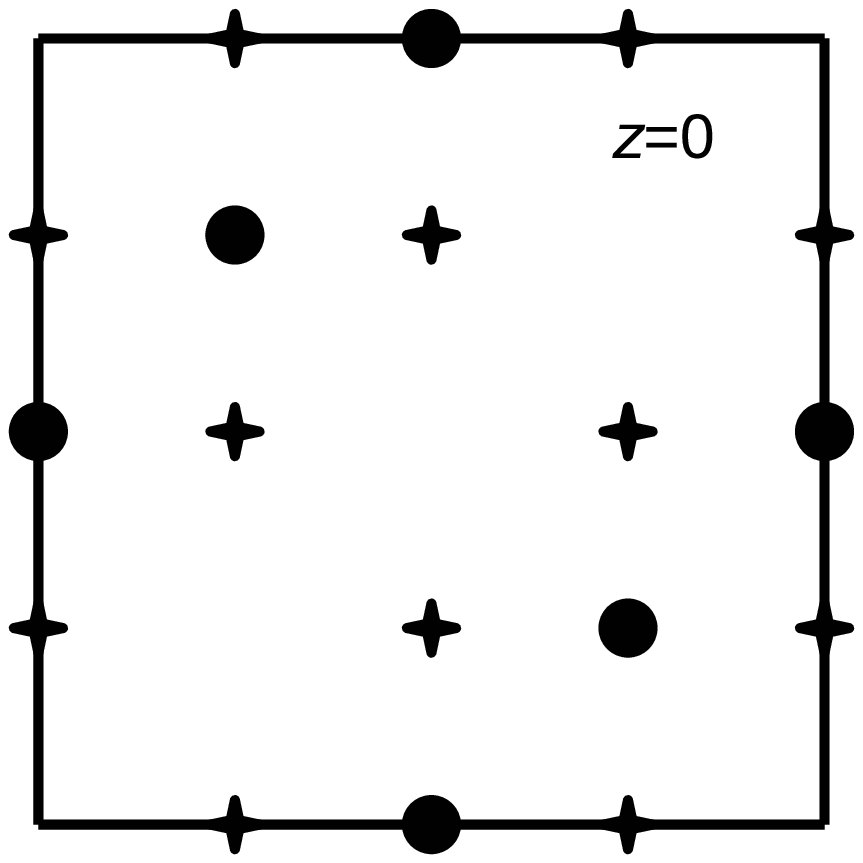}\\
\includegraphics[width=\textwidth,clip]{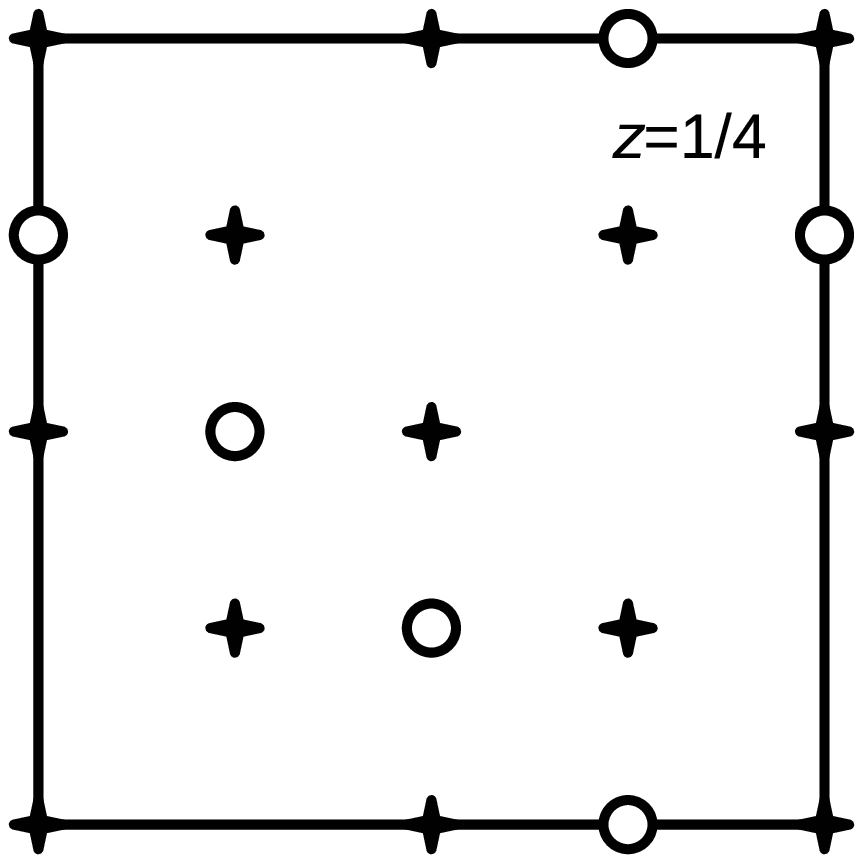}\\
\includegraphics[width=\textwidth,clip]{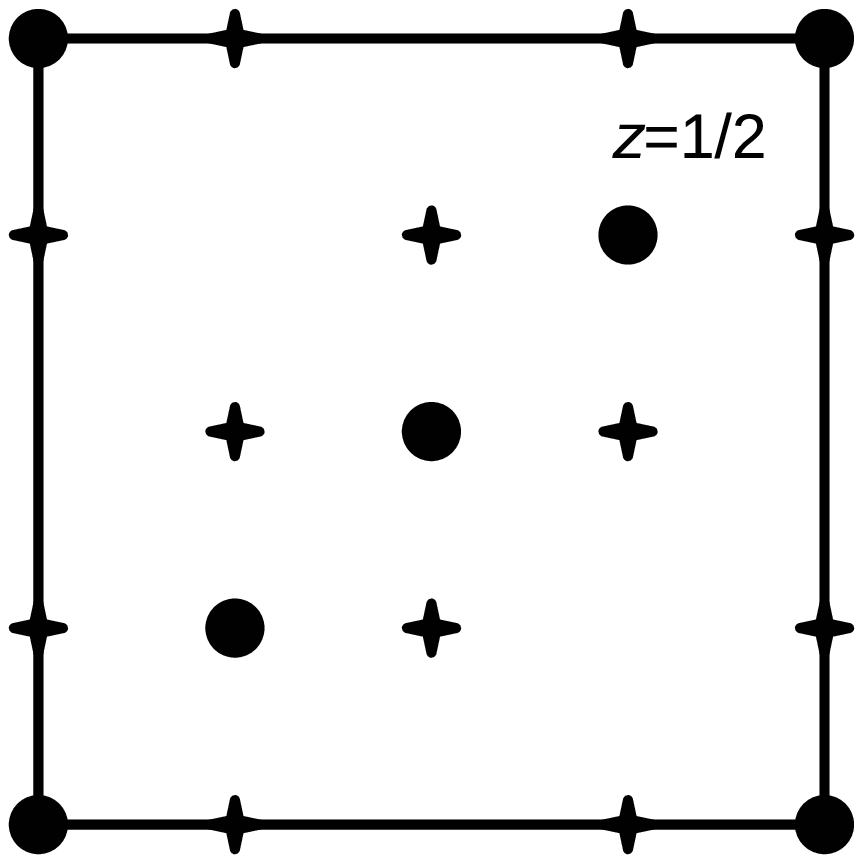}\\
\includegraphics[width=\textwidth,clip]{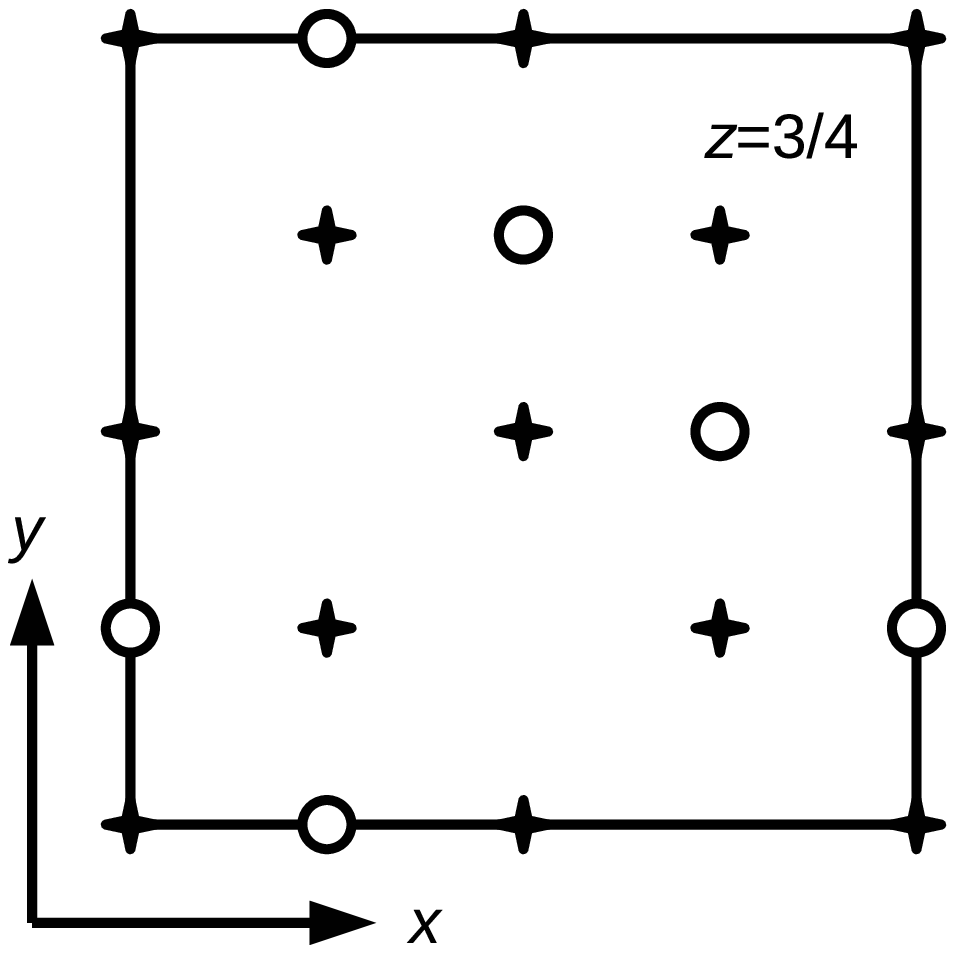}\\
\end{minipage}
}
\caption{Cation distribution within different $x$-$y$ planes for
  different values of $z$, corresponding to ``setting'' 1
  \subref{FigureInverseSpinelSetting1} and ``setting'' 3
  \subref{FigureInverseSpinelSetting3} (see main text for
  details). Only layers containing $B$ site cations are
  shown. Divalent cations (Co$^{2+}$, Ni$^{2+}$) are depicted as
  filled black circles, trivalent cations (Fe$^{3+}$) as open black
  circles, and oxygen anions as black crosses. Note that we use the
  convention where the origin is located at the midpoint between two
  $A$ sites.}
\label{FigureInverseSpinelSettings}
\end{figure}

In order to minimize the effect of this artificial symmetry lowering,
and to obtain results that are as close as possible to the average
cubic bulk symmetry seen in experiments, we apply the following
constraints during our structural relaxations. For the calculations
corresponding to the unstrained bulk case, we constrain the lattice
parameters along the three cartesian direction to be equal,
\mbox{$a=b=c$}, and we fix the $A$ site cations to their ideal cubic
positions, corresponding to \mbox{$z(\text{4e}) = \tfrac{1}{8}$}. On
the other hand, since the oxygen positions are characterized by one
free structural parameter already within cubic $Fd\bar{3}m$ symmetry
(Wyckoff position 32e $(u, u, u)$), we do not apply any constraints to
the 8i and 8h positions within $Imma$ symmetry. The relaxed bulk
structure is then found by relaxing all internal positions for
different volumes and finding the volume that minimizes the
total energy. For the relaxations corresponding to a
certain value of epitaxial strain, we constrain the two in-plane
lattice constants to be equal, and then vary the out-of-plane lattice
constant $c$ while relaxing all internal coordinates (except for
$z(\text{4e})=\tfrac{1}{8}$).

While for the unstrained bulk structure all possible $B$ site cation
distributions that can be accommodated within the primitive fcc unit
cell of the spinel structure lead to the same $Imma$ symmetry,
different cases can be distinguished once an epitaxial constraint is
applied to the resulting structure. To illustrate this, the cation
distributions for two different settings are depicted in
Figs.~\ref{FigureInverseSpinelSetting1} and
\ref{FigureInverseSpinelSetting3}. The $B$ site cation distribution
(and the oxygen positions) within different $x$-$y$ planes are shown,
corresponding to different ``heights'' $z$. The different settings
simply correspond to different orientations of the orthorhombic $Imma$
crystal axes relative to the original cubic axes.

It can be seen from Fig.~\ref{FigureInverseSpinelSettings} that the
$B$ sites are arranged in an interconnected network of chains along
$\langle 110 \rangle$-type directions. In setting 1 (Fig.~\ref{FigureInverseSpinelSetting1}), the corresponding
chains within the same $x$-$y$ plane contain alternating divalent and
trivalent cations. In contrast, for setting 3 (Fig.~\ref{FigureInverseSpinelSetting3}) each $x$-$y$ plane
contains only one unique cation species, which then alternates between
adjacent planes along the $z$ direction (see also Fig.~\ref{SpinelStructureInverse}). In setting 1 the same
alternating planes are oriented perpendicular to the $x$ direction,
whereas for setting 2 (not shown) these planes are oriented
perpendicular to the $y$ direction. If an epitaxial constraint is
applied within the $x$-$y$ plane, setting 3 becomes different from
settings 1 and 2, with the difference being the orientation of the
``substrate plane'' relative to the planes defined by the cation
order.

In view of this, we have performed all calculations corresponding to
epitaxially strained systems for both setting 1 and setting 3. The
differences between the results obtained for the two different
settings then represent a measure for the sensitivity of these results
from the specific cation arrangement used in our calculations.

\subsection{Other computational details}
\label{TheoryComputationalDetails}

All calculations presented in this work were performed using the
projector-augmented wave (PAW) method,\cite{Bloechl_PRB50_17953}
implemented in the Vienna \textit{ab initio} simulation package (VASP
4.6).\cite{Kresse_PRB47_558,Kresse_PRB49_14251,Kresse_CompMatSci6_15,Kresse_PRB54_11169}
Standard PAW potentials supplied with VASP were used in the
calculations, contributing nine valence electrons per Co
(4s$^2$3d$^7$), 16 valence electrons per Ni (3p$^6$4s$^2$3d$^8$), 14
valence electrons per Fe (3p$^6$4s$^2$3d$^6$), and 6 valence electrons
per O (2s$^2$2p$^4$). A plane wave energy cutoff of 500~eV was used,
and the Brillouin zone was sampled using different $k$-point grids
centered at the $\Gamma$ point. A 5 $\times$ 5 $\times$ 5 $k$-point
grid was used for the structural optimization and all total energy
calculations, whereas a finer 7 $\times$ 7 $\times$ 7 grid was used to
calculate densities of states (DOS). The tetrahedron method with
Bl\"{o}chl corrections was used for Brillouin zone
integration.\cite{Bloechl_PRB50_17953} We have verified that all
quantities of interest, in particular the MAEs, are well converged for
the used $k$-point grid and energy cutoff. All structural relaxations
were performed within a scalar-relativistic approximation, whereas
spin-orbit coupling was included in the calculation of the MAEs.

As already noted in Sec.~\ref{TheorySpinelStructure}, CoFe$_2$O$_4$
and NiFe$_2$O$_4$ are small gap insulators, whereas half-metallic or,
in the case of NiFe$_2$O$_4$, only marginally insulating band-structures have
been obtained in previous LSDA
calculations.\cite{Penicaud_JMagMagMat103_212} In the present work we
therefore use the LSDA+$U$ and GGA+$U$
approach,\cite{Anisimov/Zaanen/Andersen:1991} which is known to give a
good description of the electronic structure for many transition metal
oxides.\cite{Anisimov/Aryatesiawan/Liechtenstein:1997} We employ the
Hubbard ``+$U$'' correction in the simplified, rotationally invariant
version of Dudarev \textit{et al.},\cite{Dudarev_PRB57_1505} where the same value $U_\text{eff}=U-J=3$~eV is used for all
transition metal cations. The corresponding results are compared to pure GGA calculations, using the
GGA approach of Perdew, Burke, and Ernzerhof.\cite{Perdew_PRL77_3865} (We restricted the comparison to pure GGA since Perron \textit{et al.}\cite{Perron_JPCM19_346219} presented some evidence (for NiFe$_2$O$_4$) that LSDA might not be appropriate to properly describe these materials.)

Values for the local magnetic moments and atom-projected DOS
are obtained by integration of the appropriate
quantities over atom-centred spheres with radii taken from the applied PAW potentials
(1.164 \AA{} (Fe), 1.302 \AA{} (Co), and 1.058 \AA{} (Ni)), respectively.

\section{Results and discussion}
\label{ResultsAndDiscussion}

\subsection{Unstrained bulk structures}
\label{ResultsBulkStructureAndDOS}

\begin{table}
\caption{\label{TableStructuralPropertiesBulk} Optimized bulk lattice
  constants $a_0$ and bulk moduli $B$ for CoFe$_2$O$_4$ and
  NiFe$_2$O$_4$, calculated using the LSDA+$U$, GGA, and GGA+$U$
  exchange-correlation functionals in comparison with experimental
  data.}
\begin{ruledtabular}
\begin{tabular}{lcccc}
& \multicolumn{2}{c}{CoFe$_2$O$_4$} & \multicolumn{2}{c}{NiFe$_2$O$_4$} \\
& $a_0$ [\AA{}] & $B$ [GPa] & $a_0$ [\AA{}] & $B$ [GPa] \\
\hline
LSDA+$U$ & 8.231 & 206.0 & 8.196 & 213.1 \\
GGA      & 8.366 & 211.0 & 8.346 & 166.2 \\
GGA+$U$  & 8.463 & 172.3 & 8.426 & 177.1 \\
Exp. (Ref.~\onlinecite{Li_JMaterSci26_2621}) & 8.392 & 185.7 & 8.339 & 198.2 \\
Exp. (Ref.~\onlinecite{Hill_PhysChemMinerals4_317}) & 8.35 & --- &
8.325 & --- \\
\end{tabular}
\end{ruledtabular}
\end{table}

We first present our results for the unstrained bulk structures. The
calculated lattice constants and bulk moduli for both CoFe$_2$O$_4$
and NiFe$_2$O$_4$ using different exchange-correlation functionals are
summarized in Table~\ref{TableStructuralPropertiesBulk}. It can be
seen that for both CoFe$_2$O$_4$ and NiFe$_2$O$_4$ the use of LSDA+$U$
leads to an underestimation of the lattice constant and an
overestimation of the bulk modulus compared to the experimental
values, whereas the opposite is the case for GGA+$U$. The
corresponding deviations ($\sim$1-2~\% for the lattice constants) are
typical for complex transition metal oxides (see
e.g. Refs.~\onlinecite{Fennie_PRL96_205505,Neaton_PRB71_014113,Ederer_CurrOpinion9_128}).

Interestingly, the lattice constants calculated within pure GGA match
the experimental values almost perfectly for both CoFe$_2$O$_4$ and
NiFe$_2$O$_4$. However, this is somewhat fortuitous and probably due
to a cancellation of errors, as can be seen by the large discrepancies
in the bulk moduli. It will become clear in the following, that the
``+$U$'' correction is necessary in order to obtain a good description
of the electronic structure for both CoFe$_2$O$_4$ and NiFe$_2$O$_4$.

\begin{table}
\caption{\label{TableOxygenPositions} Calculated Wyckoff parameters (setting 1)
  for the oxygen anions 8h $(x, \tfrac{1}{4}, z)$ and 8i $(0, y, z)$
  for CoFe$_2$O$_4$ and NiFe$_2$O$_4$ using different
  exchange-correlation functionals. The last line lists the
  corresponding parameters resulting from Wyckoff position 32e $(u, u,
  u)$ within $Fd\bar{3}m$ symmetry, $\bar{u}$ is obtained from these relations by averaging over recalculated $u$ values for each dataset.}
\begin{ruledtabular}
\begin{tabular}{lccccc}
CoFe$_2$O$_4$ & \multicolumn{2}{c}{8i} & \multicolumn{2}{c}{8h} \\
& $x$ & $z$ & $y$ & $z$ & $\bar{u}$ \\
LDA+$U$ & 0.235 & $-$0.498 & 0.009 & $-$0.257 & 0.255 \\
GGA     & 0.240 & $-$0.496 & 0.008 & $-$0.255 & 0.255 \\
GGA+$U$ & 0.234 & $-$0.499 & 0.007 & $-$0.259 & 0.255 \\
\hline
NiFe$_2$O$_4$ & \multicolumn{2}{c}{8i} & \multicolumn{2}{c}{8h} \\
& $x$ & $z$ & $y$ & $z$ & $\bar{u}$ \\
LDA+$U$ & 0.237 & $-$0.495 & 0.010 & $-$0.258 & 0.256 \\
GGA     & 0.239 & $-$0.496 & 0.009 & $-$0.257 & 0.255 \\
GGA+$U$ & 0.235 & $-$0.496 & 0.008 & $-$0.258 & 0.256 \\
\hline
$Fd\bar{3}m$ & $\tfrac{3}{4}-2u$ & $u-\tfrac{3}{4}$ & $2u-\tfrac{1}{2}$ & $-u$ \\
\end{tabular}
\end{ruledtabular}
\end{table}

As discussed in Sec.~\ref{TheoryStructuralRelaxations} the cation
arrangement in our unit cell lowers the symmetry to orthorhombic
$Imma$, with 4 independent parameters describing the positions of the
oxygen anions at Wyckoff positions 8h and 8i, compared to one
parameter for Wyckoff position 32e in the cubic space group
$Fd\bar{3}m$. Table~\ref{TableOxygenPositions} lists the corresponding
Wyckoff parameters obtained from our structural optimizations. It can
be seen that differences between different exchange-correlation
functionals are rather small.

The last line in Table~\ref{TableOxygenPositions} indicates the
relation between the Wyckoff positions 8i and 8h in the $Imma$ space
group (setting 1) and Wyckoff position 32e $(u,u,u)$ in cubic $Fd\bar{3}m$
symmetry (assuming that no actual symmetry breaking occurs). These relations
allow us to obtain an average Wyckoff parameter $\bar{u}$, by
calculating the value of $u$ corresponding to each of the four
calculated Wyckoff parameters $x(\text{8i})$, $z(\text{8i})$,
$y(\text{8h})$, and $z(\text{8h})$ for each data set and subsequent averaging. The
resulting values for $\bar{u}$ agree very well with available experimental
data, which are 0.256 for CoFe$_2$O$_4$ and 0.257 for
NiFe$_2$O$_4$.\cite{Hill_PhysChemMinerals4_317} Furthermore, the values
calculated from the individual 8h and 8i Wyckoff parameters deviate
only very little from the average values, which indicates that the
lower symmetry used in our calculation has only a negligible effect on
the internal structural parameters.

\begin{figure}
\caption{\label{Figure_CoFe2O4_DOS_GGA_GGA+U} (Color online) Total and projected DOS per formula unit for CoFe$_2$O$_4$. Left (right) panels correspond to GGA (GGA+$U$) calculations. The $d$ states of Co($O_h$) (upper panels), Fe($O_h$) (middle panels), and Fe($T_d$) (lower panels) are separated into $t_{2g}$ (green/dark grey) and $e_g$ (red/black) contributions for the $O_h$ sites and into $e$ (green/dark grey) and $t_2$ (red/black) contributions for the $T_d$ sites. The total DOS is shown as shaded grey area in all panels. Majority (minority) spin projections correspond to positive (negative) values.}
\includegraphics[width=0.5\textwidth,clip]{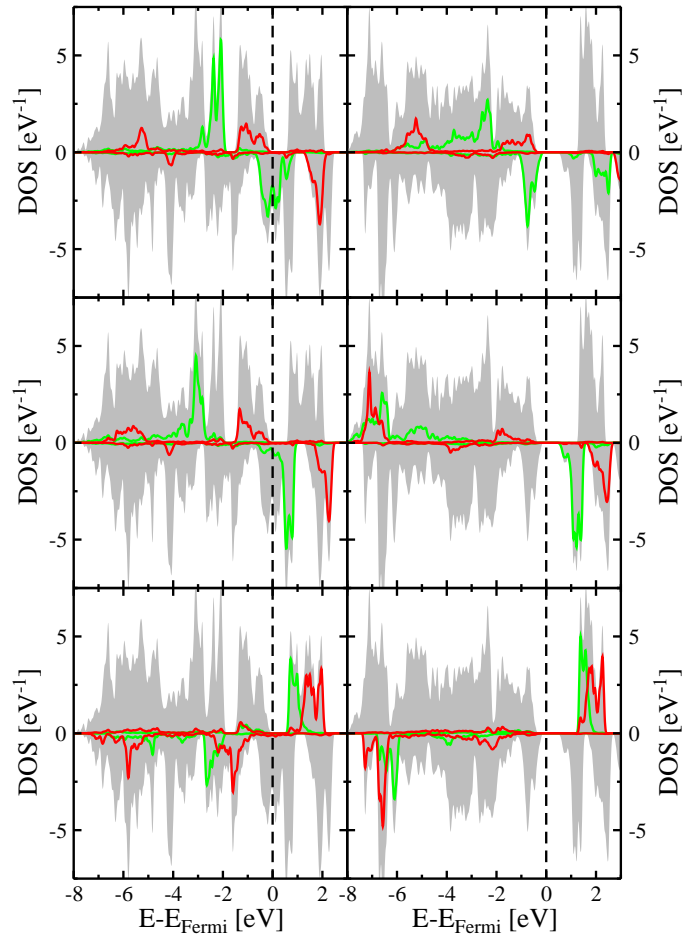}
\end{figure}

\begin{figure}
\caption{\label{Figure_NiFe2O4_DOS_GGA_GGA+U} (Color online) Total and projected DOS per formula unit for NiFe$_2$O$_4$. Left (right) panels correspond to GGA (GGA+$U$) calculations. The $d$ states of Co($O_h$) (upper panels), Fe($O_h$) (middle panels), and Fe($T_d$) (lower panels) are separated into $t_{2g}$ (green/dark grey) and $e_g$ (red/black) contributions for the $O_h$ sites and into $e$ (green/dark grey) and $t_2$ (red/black) contributions for the $T_d$ sites. The total DOS is shown as shaded grey area in all panels. Majority (minority) spin projections correspond to positive (negative) values.}
\includegraphics[width=0.5\textwidth,clip]{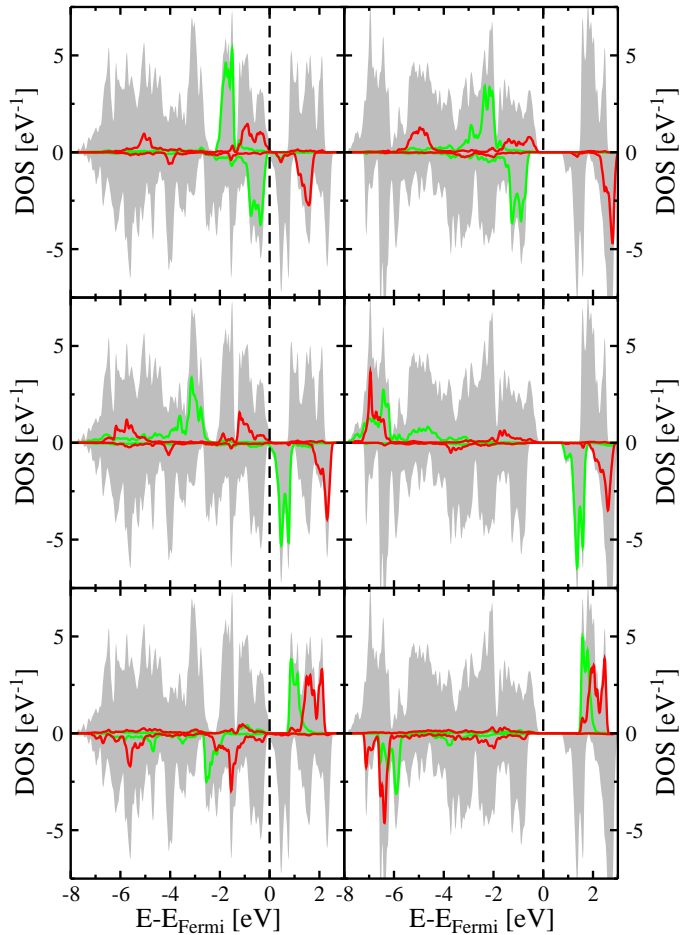}
\end{figure}

Figs.~\ref{Figure_CoFe2O4_DOS_GGA_GGA+U} and
\ref{Figure_NiFe2O4_DOS_GGA_GGA+U} show the calculated spin-decomposed
densities of states (DOS) for CoFe$_2$O$_4$ and NiFe$_2$O$_4$,
respectively, within the optimized bulk structure and using both pure
GGA and GGA+$U$. Both the total DOS per formula unit and the projected DOS per ion for the $d$
states of the various transition metal cations are shown, the latter
separated into $t_{2g}$/$e_g$ and $e$/$t_2$ contributions,
respectively. The DOS calculated within LSDA+$U$ (not shown) do not show any
significant differences compared to the ones calculated using GGA+$U$.

From the projected DOS it can be seen that all transition metal
cations are in high spin states, with one spin-projection completely
occupied, and that the cubic component of the crystal field on the
octahedrally-coordinated ($O_h$) sites lowers the $t_{2g}$ states
relative to the $e_g$ states, whereas on the tetrahedrally-coordinated
($T_d$) sites the $e$ states are slightly lower in energy than the
$t_2$ states. Due to the N{\'e}el-type ferrimagnetic order, the local
majority spin direction on the $T_d$ sites is reversed relative to the
$O_h$ sites.

It is apparent that within GGA CoFe$_2$O$_4$ turns out to be a
half-metal, in contrast to the insulating behavior found in
experiment.\cite{Jonker_JPhysChemSol9_165} This is similar to what
has been found in previous LSDA
calculations.\cite{Penicaud_JMagMagMat103_212} The half-metallicity is
due to the partial filling of the minority $t_{2g}$ states of
Co(O$_h$), which in turn results from the formal $d^7$ configuration
of the Co$^{2+}$ cation (see upper left panel of
Fig.~\ref{Figure_CoFe2O4_DOS_GGA_GGA+U}). This apparent deficiency of
the GGA approach is corrected within the GGA+$U$ calculation as can be
seen in the right part of Fig.~\ref{Figure_CoFe2O4_DOS_GGA_GGA+U}. We
note that this is very similar to the case of rocksalt CoO, which also
contains Co$^{2+}$ with a $d^7$ electron configuration that leads to a
metallic solution in pure LSDA,\cite{Terakura_PRB30_4734} whereas
application of the DFT+$U$ approach leads to an insulating state in
agreement with the experimental
observations.\cite{Anisimov/Zaanen/Andersen:1991} 

The Hubbard correction splits the occupied and unoccupied parts of the
minority spin $t_{2g}$ states of the Co$^{2+}$ cations (see upper
right panel of Fig.~\ref{Figure_CoFe2O4_DOS_GGA_GGA+U}), thereby
opening an energy gap. In addition, the local spin splitting on the Fe
cation is drastically enhanced, shifting the local majority spin $d$
states towards the bottom of the valence band. We note that once the
value of $U_\text{eff}$ on the Co sites is large enough to push the
corresponding unoccupied minority spin $t_{2g}$ states above the
lowest minority-spin Fe($O_h$) states, the width of the band gap is
determined by the difference in energy between these lowest unoccupied
minority-spin Fe states and the highest occupied minority spin
$t_{2g}$ states of Co($O_h$). Therefore, a further increase of
$U_\text{eff}$ on the Co sites does not significantly change the size
of the band gap. Similarly, the band gap depends only weakly on the
specific value of $U_\text{eff}$ on the Fe sites.

The gap size of 0.9~eV obtained within GGA+$U$ for the chosen values
of $U_\text{eff}$ is comparable to the 0.63~eV obtained by Antonov
\textit{et al.} using LSDA+$U$ with $U_\text{eff}$=4.0 eV for the
Co(O$_h$) and $U_\text{eff}$=4.5 eV for the Fe(O$_h$) and Fe(T$_d$)
cations,\cite{Antonov_PRB67_024417} and also agrees well with the
value of 0.8~eV reported by Szotek \textit{et al.} utilizing a
self-interaction corrected LSDA approach.\cite{Szotek_PRB74_174431}

Ni$^{2+}$ formally has one additional electron compared to Co$^{2+}$,
leading to a fully occupied minority spin $t_{2g}$ manifold for the
Ni$^{2+}$ cation within a cubic ($O_h$) crystal field. Accordingly,
NiFe$_2$O$_4$ exhibits a tiny gap of $\sim$0.1~eV between the occupied
minority spin $t_{2g}$ states of Ni($O_h$) and the unoccupied minority
$t_{2g}$ states of Fe(O$_h$) even in pure GGA (see left panels of
Fig.~\ref{Figure_NiFe2O4_DOS_GGA_GGA+U}). However, the use of GGA+$U$
leads to a significant enlargement of this energy gap to a more
realistic value of 0.97~eV for the chosen values of
$U_\text{eff}$. This is in good agreement with band gaps of 0.99~eV
and 0.98~eV reported by Antonov \textit{et
  al.}~\cite{Antonov_PRB67_024417} and Szotek \textit{et
  al.},~\cite{Szotek_PRB74_174431} respectively. Similar to the case
of CoFe$_2$O$_4$ the Hubbard correction also leads to a strong
enhancement of the local spin splitting on the Fe sites in
NiFe$_2$O$_4$.

\begin{table}
\caption{\label{TableMagneticMomentsBulk} Calculated magnetic moments
  (in $\mu_\text{B}$) for bulk CoFe$_2$O$_4$ and NiFe$_2$O$_4$ using
  different exchange-correlation functionals. The total magnetic
  moment per formula unit amounts to 3 $\mu_\text{B}$ (2
  $\mu_\text{B}$) for CoFe$_2$O$_4$ (NiFe$_2$O$_4$), respectively.}
\begin{ruledtabular}
\begin{tabular}{lccc}
CoFe$_2$O$_4$ & Co($O_{h}$) & Fe($O_{h}$) & Fe($T_{d}$) \\
\hline
LSDA+U      & +2.52 & +3.99 & $-$3.82 \\
GGA         & +2.43 & +3.66 & $-$3.45 \\
GGA+U       & +2.62 & +4.10 & $-$3.98 \\
\hline
NiFe$_2$O$_4$ & Ni($O_{h}$) & Fe($O_{h}$) & Fe($T_{d}$) \\
\hline
LSDA+U      & +1.49 & +4.00 & $-$3.82 \\
GGA         & +1.36 & +3.71 & $-$3.46 \\
GGA+U       & +1.58 & +4.11 & $-$3.97 \\
\end{tabular}
\end{ruledtabular}
\end{table}

Table~\ref{TableMagneticMomentsBulk} shows the local magnetic moments
of the transition metal cations per formula unit calculated within the
three different approaches. The total magnetic moment is independent
of the applied exchange-correlation functional, and equal to the
integer value that follows from the formal electron configuration of
the transition metal cations and the N{\'e}el-type ferrimagnetic
arrangement (3 $\mu_B$ for CoFe$_2$O$_4$ and 2 $\mu_B$ for
NiFe$_2$O$_4$). This is a result of the either half-metallic or
insulating character of the underlying electronic
structures. Nevertheless, as can be seen in
Table~\ref{TableMagneticMomentsBulk}, the Hubbard correction has a
significant influence on the spatial distribution of the magnetization
density and the use of GGA+$U$ (and LSDA+$U$) leads to more localized
magnetic moments compared to GGA, indicated by the increased magnetic
moments corresponding to the different cation sites.

The results presented in this section indicate that for a realistic
and consistent description of the structural, electronic, and magnetic
properties of both CoFe$_2$O$_4$ and NiFe$_2$O$_4$, a Hubbard
correction to either LSDA or GGA is required. In the following we will
therefore present only results obtained within the LSDA+$U$ and
GGA+$U$ approaches.

\subsection{Epitaxial strain and elastic properties}
\label{ResultsStrainAndElasticProperties}

\begin{figure}
\includegraphics[width=0.45\textwidth,clip]{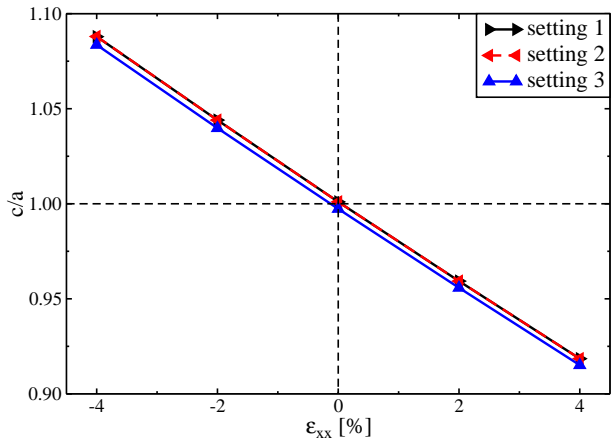}
\caption{\label{Figure_NiFe2O4_GGA+U_coa} (Color online) Calculated $c/a$ ratio of
  NiFe$_{2}$O$_{4}$ as a function of epitaxial strain
  $\varepsilon_{xx}$ obtained from GGA+$U$ calculations and different
  cation arrangements (``settings'') on the $O_h$ sites.}
\end{figure}

Fig.~\ref{Figure_NiFe2O4_GGA+U_coa} shows the relaxed $c/a$ ratio of
NiFe$_2$O$_4$ as function of the epitaxial constraint
$\varepsilon_{xx}$, obtained from GGA+$U$ calculations as described in
Sec.~\ref{TheoryStructuralRelaxations}. The case of CoFe$_2$O$_4$ is
very similar. Two important things can be seen from this. First, due
to the orthorhombic $Imma$ symmetry of the chosen cation arrangement
on the $O_h$ sites, the $c/a$ ratio is not exactly equal to 1 at zero
strain. However, this effect is clearly negligible compared to the
changes in $c/a$ induced by epitaxial strains of order $\sim$1~\% and
therefore does not affect our further analysis. Second, the slope of
the $c/a$ ratio, which characterizes the elastic response of the
material, is nearly completely unaffected by the different cation
arrangements.

From the data shown in Fig.~\ref{Figure_NiFe2O4_GGA+U_coa} we can
therefore obtain the 2-dimensional Poisson ratio $\nu_\text{2D}$
(Eq.~\eqref{Equation2DPoissonRatio}), which relates in-plane and
out-of-plane strains. Together with the bulk moduli listed in
Table~\ref{TableStructuralPropertiesBulk}, we can then determine the
two elastic constants $C_{11}$ and $C_{12}$ from
Eqs.~\eqref{EquationBulkModulus} and \eqref{Equation2DPoissonRatio}.

\begin{table}
\caption{\label{TableElasticProperties} Bulk modulus $B$,
  2-dimensional Poisson ratio $\nu_{2D}$, and elastic coefficients
  $C_{11}$ and $C_{12}$ for CoFe$_{2}$O$_{4}$ and NiFe$_{2}$O$_{4}$,
  obtained for different exchange-correlation potentials and different cation arrangements (``setting'' $s$), in comparison to
  experimental data. The experimental $\nu_{2D}$ has been evaluated
  from Eq.~\eqref{Equation2DPoissonRatio} using the experimental
  elastic constants.}
\begin{ruledtabular}
\begin{tabular}{lccccc}
CoFe$_{2}$O$_{4}$ & $s$ & $B$ [GPa] & $\nu_{2D}$ & $C_{11}$ [GPa] & $C_{12}$ [GPa] \\
\hline
LSDA+$U$          & 1   & 206.0     & 1.191      & 282.0          & 167.9 \\
                  & 3   &           & 1.185      & 282.7          & 167.6 \\
GGA+$U$           & 1   & 172.3     & 1.147      & 240.8          & 138.1 \\
                  & 3   &           & 1.132      & 242.5          & 137.3 \\
Exp.\cite{Li_JMaterSci26_2621} & & 185.7 & 1.167 & 257.1 & 150.0 \\
\hline
NiFe$_{2}$O$_{4}$ & $S$ & $B$ [GPa] & $\nu_{2D}$ & $C_{11}$ [GPa] & $C_{12}$ [GPa] \\
\hline
LSDA+$U$          & 1   & 213.1     & 1.172      & 294.4          & 172.5 \\
                  & 3   &           & 1.167      & 295.1          & 172.2 \\
GGA+$U$           & 1   & 177.1     & 1.115      & 251.2          & 140.0 \\
                  & 3   &           & 1.106      & 252.2          & 139.5 \\
Exp.\cite{Li_JMaterSci26_2621} & & 198.2 & 1.177 & 273.1 & 160.7 \\
\end{tabular}
\end{ruledtabular}
\end{table}

The calculated 2-dimensional Poisson ratios $\nu_{2D}$ and elastic
constants $C_{11}$ and $C_{12}$, together with the bulk moduli already
presented in Table~\ref{TableStructuralPropertiesBulk}, are listed in
Table~\ref{TableElasticProperties}, and are compared with experimental
results from Ref.~\onlinecite{Li_JMaterSci26_2621}. It can be seen
that, as already pointed out, the specific cation arrangement has
nearly no influence on the value of $\nu_{2D}$ and thus $C_{11}$ and
$C_{12}$. On the other hand, the specific choice of either LSDA+$U$ or
GGA+$U$ has a noticeable effect. Similar to the case of the bulk
modulus, we observe an overestimation (underestimation) of the elastic
constants $C_{11}$ and $C_{12}$ in the LSDA+$U$ (GGA+$U$)
calculations. The same general trend holds for the 2-dimensional
Poisson-ratio of CoFe$_{2}$O$_{4}$, while for NiFe$_{2}$O$_{4}$ the
use of LSDA+$U$ also slightly underestimates $\nu_{2D}$. Overall the
deviations are only within a few percent of the experimental data
(1-2~\% for CoFe$_{2}$O$_{4}$ and up to 6~\% for NiFe$_{2}$O$_{4}$),
and we therefore conclude that both LSDA+$U$ and GGA+$U$ allow for a
good description of the strain response of CoFe$_{2}$O$_{4}$ and
NiFe$_{2}$O$_{4}$.

\subsection{Magnetoelastic coupling}
\label{ResultsMAEAndMagnetoelasticProperties}

\begin{figure}
\caption{\label{FigureCoFe2O4StrainMAE} (Color online) CoFe$_2$O$_4$ GGA+$U$ total
  energy differences for orientation of the magnetization along
  various crystallographic in-plane directions with respect to the
  [001] direction (black solid lines): $\blacktriangle$ [100],
  $\blacktriangledown$ [010], $\blacktriangleleft$ [110], and
  $\blacktriangleright$ [1$\bar{1}$0]. Red broken lines denote
  crystallographic directions which include also out-of-plane
  components, namely [101] (dashed-dotted line), [011]
  (dashed-double dotted line), and [111] (dashed
  line). Left (right) panels contain the results corresponding to setting 1 (setting 3).}
\includegraphics[width=0.475\textwidth,clip]{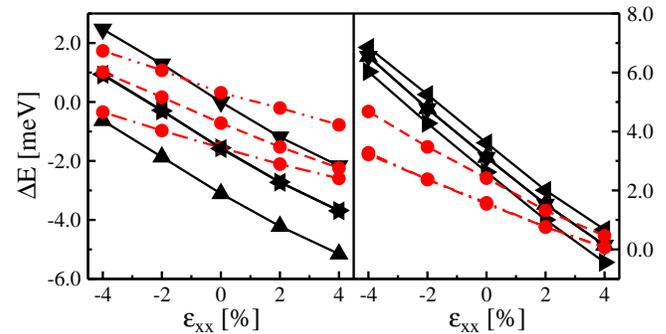}
\end{figure}

\begin{figure}
\caption{\label{FigureNiFe2O4StrainMAE} (Color online) NiFe$_2$O$_4$ LSDA+$U$
  (GGA+$U$) total energy differences for orientation of the magnetization along
  various crystallographic in-plane directions with respect to the
  [001] direction are shown in the upper (lower) panels (black solid lines): $\blacktriangle$ [100],
  $\blacktriangledown$ [010], $\blacktriangleleft$ [110], and
  $\blacktriangleright$ [1$\bar{1}$0]. Red broken lines denote
  crystallographic directions which include also out-of-plane
  components, namely [101] (dashed-dotted line), [011]
  (dashed-double dotted line), and [111] (dashed
  line). Left (right) panels contain the results corresponding to setting 1 (setting 3).}
\includegraphics[width=0.475\textwidth,clip]{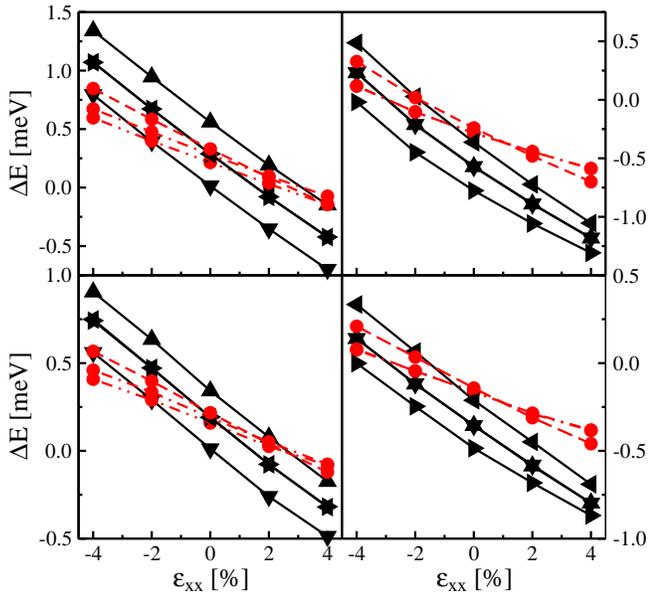}
\end{figure}

The calculated MAEs, defined as the energy differences for various
orientations of the magnetization relative to the energy for
orientation of the magnetization parallel to the [001] direction, are
depicted in Fig.~\ref{FigureCoFe2O4StrainMAE} for CoFe$_2$O$_4$
(GGA+$U$ only) and in Fig.~\ref{FigureNiFe2O4StrainMAE} for
NiFe$_2$O$_4$ (both LSDA+$U$ and GGA+$U$). It can be seen that the
calculated MAEs for CoFe$_2$O$_4$ are roughly five to
six times larger than in NiFe$_2$O$_4$. Furthermore, to a good approximation,
the calculated energy differences exhibit a linear dependence on
strain. Deviations from this linear behavior are most pronounced for
the case of NiFe$_2$O$_4$ within setting 3. Since the pure elastic
response shown in Fig.~\ref{Figure_NiFe2O4_GGA+U_coa} does not exhibit
any significant non-linearities, we conclude that higher order
magnetoelastic terms are responsible for the slightly non-linear
behavior of the MAE in this case.

It can also be seen, that in all cases the strain dependence, i.e. the
slope of the various curves shown in
Figs.~\ref{FigureCoFe2O4StrainMAE} and \ref{FigureNiFe2O4StrainMAE},
is largest for the in-plane versus out-of-plane energy differences,
i.e. for orientation of the magnetization along the [100], [010],
[110], and [1$\bar{1}$0] directions (compared to the [001] direction),
consistent with Eqs.~\eqref{EquationB1}. Thus, tensile strain
favors perpendicular anisotropy (easy axis perpendicular to the
``substrate'') and compressive strain favors in-plane orientation of
the magnetization, i.e. $B_1 > 0$. For sufficient amount of strain the
easy axis of magnetization will therefore always be oriented either
in-plane or out-of-plane, consistent with various experimental
observations in thin CoFe$_2$O$_4$ films under tensile
strain.~\cite{Lisfi_JPCM17_1399,Lisfi_PRB76_054405,Gao_JPhysD42_175006}

According to the phenomenological magnetoelastic theory for a cubic
crystal discussed in Sec.~\ref{TheoryMagnetoelasticEnergyAndMAE}, in
particular Eqs.~\eqref{EquationB1}, the strain dependence of
the in-plane versus out-of-plane anisotropy should be stronger by a
factor of 2 compared to the anisotropy corresponding to [101] or [011]
orientation of the magnetization, and by a factor of 3/2 compared to
[111] orientation. We note that these ratios are very well observed by
the calculated anisotropy energies shown in
Figs.~\ref{FigureCoFe2O4StrainMAE} and
\ref{FigureNiFe2O4StrainMAE}. This indicates that the
strain-dependence of the calculated anisotropy energies is rather
independent of the specific cation arrangement, which allows us to
obtain the magnetoelastic constants of CoFe$_2$O$_4$ and NiFe$_2$O$_4$ from our
calculations.

On the other hand, we recall that due to the specific cation
arrangement used in the calculations, and the resulting symmetry
lowering from cubic to orthorhombic, the three cubic axes are not
equivalent even for zero strain. For setting 1 the $y$ and $z$
directions are equivalent, but different from $x$, whereas for setting
3 $x$ and $y$ are equivalent but different from $z$. This is reflected
in the calculated anisotropy energies for zero strain
($\varepsilon_{xx}=0$), which in setting 1 are largest between $z$/$y$
and the $x$ direction. (Note that for zero strain both settings are
completely equivalent apart from a rotation of the coordinate axes by
120$^\circ$ around the [111] direction.) The anisotropy induced by
the symmetry-lowering to orthorhombic is therefore at least of the
same magnitude as the cubic anisotropy in the disordered inverse
spinel structure, and the calculated anisotropy energies for zero
strain are therefore not representative for the inverse spinel
structure with random distribution of cations on the $O_h$ sites,
i.e. the cubic anisotropy constant $K$ cannot be determined from our
calculations.

From the data shown on Figs.~\ref{FigureCoFe2O4StrainMAE} and
\ref{FigureNiFe2O4StrainMAE}, we thus obtain the magnetoelastic
coefficients $B_1$ by using the appropriate equation out of
Eqs.~\eqref{EquationB1} (together with the calculated values for
$\nu_\text{2D}$) for each calculated energy difference individually,
and then average over the resulting values for $B_1$ within the same
setting and for the same exchange-correlation functional. From the
so-obtained magnetoelastic coefficients $B_1$ we then calculate the
linear magnetostriction coefficient $\lambda_{100}$ from
Eq.~\eqref{EquationLambda100} using the elastic moduli determined in
Sec.~\ref{ResultsStrainAndElasticProperties}. The resulting values for
both $B_1$ and $\lambda_{100}$ are listed in
Table~\ref{TableMagneticProperties}.

\begin{table}
\caption{\label{TableMagneticProperties}Magnetoelastic coupling
  constant $B_1$ and magnetostriction constant $\lambda_{100}$ for
  CoFe$_2$O$_4$ and NiFe$_2$O$_4$ using different exchange-correlation
  functionals and cation arrangements (``setting'' $s$) in comparison with available
  experimental data.}
\begin{ruledtabular}
\begin{tabular}{lccccc}
& & \multicolumn{2}{c}{CoFe$_{2}$O$_{4}$} & \multicolumn{2}{c}{NiFe$_{2}$O$_{4}$} \\
& $s$ & B$_{1}$ & $\lambda_{100}$    & B$_{1}$ & $\lambda_{100}$    \\
& & [MPa]   & ($\times 10^{-6}$) & [MPa]   & ($\times 10^{-6}$) \\
\hline
LSDA+$U$ & 1 &  ---  & ---    & 10.03 & $-$54.9 \\
         & 3 &  ---  & ---    &  9.65 & $-$52.3 \\
GGA+$U$  & 1 & 29.23 & $-$189.7 &  6.72 & $-$40.3 \\
         & 3 & 39.74 & $-$251.7 &  6.08 & $-$35.9 \\
\hline
Exp.     &  &      & $-$225.0\footnotemark[1] &          & $-$50.9\footnotemark[4] \\
         &  &      & $-$250.0\footnotemark[2] &          & $-$36.0\footnotemark[5] \\
         &  &      & $-$590.0\footnotemark[3]
\end{tabular}
\end{ruledtabular}
\footnotetext[1]{Polycrystalline CoFe$_2$O$_4$
  (Ref.~\onlinecite{Chen_IEEETransMagn35_3652}).}
\footnotetext[2]{Single crystals with Co$_{1.1}$Fe$_{1.9}$O$_4$
  composition (Ref.~\onlinecite{Bozorth_PhysRev99_1788}).}
\footnotetext[3]{Single crystals with Co$_{0.8}$Fe$_{2.2}$O$_4$
  composition (Ref.~\onlinecite{Bozorth_PhysRev99_1788}).}
\footnotetext[4]{Single crystals of NiFe$_2$O$_4$
  (Ref.~\onlinecite{Smith_JAP37_1001}).}  
\footnotetext[5]{Single crystals with Ni$_{0.8}$Fe$_{2.2}$O$_4$
  composition (Ref.~\onlinecite{Bozorth_PhysRev99_1788}).}
\end{table}

It can be seen that overall the calculated magnetostriction constants
are in very good agreement with available experimental data, despite
the difficulties related to the specific choice of cation arrangement
and exchange-correlation functional. In particular the large
difference in magnetostriction between CoFe$_2$O$_4$ and NiFe$_2$O$_4$ is well
reproduced by the calculations. For NiFe$_2$O$_4$ the difference in the
calculated values for $\lambda_{100}$ due to the use of either
LSDA+$U$ and GGA+$U$ is larger than the effect of the different cation
settings. For CoFe$_2$O$_4$ the cation arrangement seems to have a larger
influence on $\lambda_{100}$ than for NiFe$_2$O$_4$. This is consistent
with the large spread in the experimentally obtained magnetostriction
for CoFe$_2$O$_4$, where different preparation conditions can lead to
differences in cation distribution/inversion or slightly
off-stoichiometric compositions. In addition, only very few
measurements have been performed on single crystals, whereas for
polycrystalline samples only a superposition of $\lambda_{100}$ and
$\lambda_{111}$ is measured.

The rather good agreement between our calculated values for
$\lambda_{100}$ and the available experimental data demonstrates that
in principle a quantitative calculation of magnetostriction in spinel
ferrites is feasible, in spite of the difficulties related to the
inverse spinel structure with random cation distribution on the $B$
site, and the usual difficulties regarding an accurate description of
exchange and correlation effects in transition metal oxides.

\section{Summary and outlook}
\label{SummaryAndOutlook}

In summary, we have presented a systematic first principles study of
the effect of epitaxial strain on the structural and
magneto-structural properties of CoFe$_2$O$_4$ and NiFe$_2$O$_4$
spinel ferrites. Special care was taken to assess the quantitative
uncertainties resulting from different treatments of
exchange-correlation effects and different cation arrangements used to
represent the inverse spinel structure.

It has been shown, in agreement with earlier works, that ``beyond
LSDA/GGA'' methods are required for a proper description of the
electronic and magnetic properties of spinel ferrites CoFe$_2$O$_4$ and NiFe$_2$O$_4$.
The ``+$U$'' approach used in the present work leads to a realistic
electronic structure and good quantitative agreement with available
experimental data for lattice, elastic, and magnetostrictive
constants.

We find that the specific cation arrangement used to represent the
inverse spinel structure has only little effect on the structural
properties. The corresponding effect on the magnetoelastic constants
is also weak, with a somewhat stronger influence in the case of
CoFe$_2$O$_4$. The latter fact is consistent with the considerable spread in
the reported values for the magnetostriction constant for different
samples of this material. In general, the starting point for the
``+$U$'' correction, i.e. either LSDA or GGA, has a somewhat stronger
influence on the calculated materials constants than the different
cation settings.

However, in spite of these uncertainties, the overall agreement
between our results and experimental data is very good. In particular,
the calculated magnetostriction constants $\lambda_{100}$ of about
$-45$~ppm for NiFe$_2$O$_4$ and about $-220$~ppm for CoFe$_2$O$_4$
fall well into the spectra of available experimental values obtained
from different samples (see
Table~\ref{TableMagneticProperties}). Consistent with the negative
sign of $\lambda_{100}$, the easy magnetization direction changes from
in-plane for compressive epitaxial strain to out-of-plane for tensile
strain. This gives further confirmation that the reorientation of the
easy axis observed experimentally in thin films of CoFe$_2$O$_4$ under
different conditions is indeed predominantly
strain-driven.\cite{Lisfi_JPCM17_1399,Lisfi_PRB76_054405,Gao_JPhysD42_175006}

In summary, our results indicate that a quantitative description of
both structural and magnetoelastic properties in spinel ferrites is
possible within the DFT+$U$ approach, which opens the way for future
computational studies of these materials. Such calculations can then provide important information regarding the
effect of cation inversion and off-stoichiometry, which can be used to
optimize magnetostriction constants and anisotropy in spinel ferrites.

\begin{acknowledgments}
This work was supported by Science Foundation Ireland under
Ref.~SFI-07/YI2/I1051 and made use of computational facilities
provided by the Trinity Centre for High Performance Computing (TCHPC)
and the Irish Centre for High-End Computing (ICHEC). D.F. acknowledges
fruitful discussions with M. Richter and K. Koepernik.
\end{acknowledgments}

\bibliography{references}

\begin{thebibliography}{51}
\expandafter\ifx\csname natexlab\endcsname\relax\def\natexlab#1{#1}\fi
\expandafter\ifx\csname bibnamefont\endcsname\relax
  \def\bibnamefont#1{#1}\fi
\expandafter\ifx\csname bibfnamefont\endcsname\relax
  \def\bibfnamefont#1{#1}\fi
\expandafter\ifx\csname citenamefont\endcsname\relax
  \def\citenamefont#1{#1}\fi
\expandafter\ifx\csname url\endcsname\relax
  \def\url#1{\texttt{#1}}\fi
\expandafter\ifx\csname urlprefix\endcsname\relax\def\urlprefix{URL }\fi
\providecommand{\bibinfo}[2]{#2}
\providecommand{\eprint}[2][]{\url{#2}}

\bibitem[{\citenamefont{Brabers}(1995)}]{Brabers1995189}
\bibinfo{author}{\bibfnamefont{V.~A.~M.} \bibnamefont{Brabers}}
  (\bibinfo{publisher}{Elsevier}, \bibinfo{year}{1995}),
  vol.~\bibinfo{volume}{8} of \emph{\bibinfo{series}{Handbook of Magnetic
  Materials}}, pp. \bibinfo{pages}{189 -- 324}.

\bibitem[{\citenamefont{Zheng et~al.}(2004)\citenamefont{Zheng, Wang, Loﬂand,
  Ma, Mohaddes-Ardabili, Zhao, Salamanca-Riba, Shinde, Ogale, Bai
  et~al.}}]{Zheng_Science303_661}
\bibinfo{author}{\bibfnamefont{H.}~\bibnamefont{Zheng}},
  \bibinfo{author}{\bibfnamefont{J.}~\bibnamefont{Wang}},
  \bibinfo{author}{\bibfnamefont{S.~E.} \bibnamefont{Loﬂand}},
  \bibinfo{author}{\bibfnamefont{Z.}~\bibnamefont{Ma}},
  \bibinfo{author}{\bibfnamefont{L.}~\bibnamefont{Mohaddes-Ardabili}},
  \bibinfo{author}{\bibfnamefont{T.}~\bibnamefont{Zhao}},
  \bibinfo{author}{\bibfnamefont{L.}~\bibnamefont{Salamanca-Riba}},
  \bibinfo{author}{\bibfnamefont{S.~R.} \bibnamefont{Shinde}},
  \bibinfo{author}{\bibfnamefont{S.~B.} \bibnamefont{Ogale}},
  \bibinfo{author}{\bibfnamefont{F.}~\bibnamefont{Bai}}, \bibnamefont{et~al.},
  \bibinfo{journal}{Science} \textbf{\bibinfo{volume}{303}},
  \bibinfo{pages}{661} (\bibinfo{year}{2004}).

\bibitem[{\citenamefont{Zavaliche et~al.}(2005)\citenamefont{Zavaliche, Zheng,
  Mohaddes-Ardabili, Yang, Zhan, Shafer, Reilly, Chopdekar, Jia, Wright
  et~al.}}]{Zavaliche_et_al:2005}
\bibinfo{author}{\bibfnamefont{F.}~\bibnamefont{Zavaliche}},
  \bibinfo{author}{\bibfnamefont{H.}~\bibnamefont{Zheng}},
  \bibinfo{author}{\bibfnamefont{L.}~\bibnamefont{Mohaddes-Ardabili}},
  \bibinfo{author}{\bibfnamefont{S.~Y.} \bibnamefont{Yang}},
  \bibinfo{author}{\bibfnamefont{Q.}~\bibnamefont{Zhan}},
  \bibinfo{author}{\bibfnamefont{P.}~\bibnamefont{Shafer}},
  \bibinfo{author}{\bibfnamefont{E.}~\bibnamefont{Reilly}},
  \bibinfo{author}{\bibfnamefont{R.}~\bibnamefont{Chopdekar}},
  \bibinfo{author}{\bibfnamefont{Y.}~\bibnamefont{Jia}},
  \bibinfo{author}{\bibfnamefont{P.}~\bibnamefont{Wright}},
  \bibnamefont{et~al.}, \bibinfo{journal}{Nano Letters}
  \textbf{\bibinfo{volume}{5}}, \bibinfo{pages}{1793} (\bibinfo{year}{2005}).

\bibitem[{\citenamefont{Dix et~al.}(2007)\citenamefont{Dix, Skumryev, Laukhin,
  F\`{a}brega, S\'{a}nchez, and Fontcuberta}}]{Dix_MatSciEngB144_127}
\bibinfo{author}{\bibfnamefont{N.}~\bibnamefont{Dix}},
  \bibinfo{author}{\bibfnamefont{V.}~\bibnamefont{Skumryev}},
  \bibinfo{author}{\bibfnamefont{V.}~\bibnamefont{Laukhin}},
  \bibinfo{author}{\bibfnamefont{L.}~\bibnamefont{F\`{a}brega}},
  \bibinfo{author}{\bibfnamefont{F.}~\bibnamefont{S\'{a}nchez}},
  \bibnamefont{and}
  \bibinfo{author}{\bibfnamefont{J.}~\bibnamefont{Fontcuberta}},
  \bibinfo{journal}{Materials Science and Engineering: B}
  \textbf{\bibinfo{volume}{144}}, \bibinfo{pages}{127} (\bibinfo{year}{2007}).

\bibitem[{\citenamefont{Muralidharan et~al.}(2008)\citenamefont{Muralidharan,
  Dix, Skumryev, Varela, S\'{a}nchez, and
  Fontcuberta}}]{Muralidharan_JAP103_07E301}
\bibinfo{author}{\bibfnamefont{R.}~\bibnamefont{Muralidharan}},
  \bibinfo{author}{\bibfnamefont{N.}~\bibnamefont{Dix}},
  \bibinfo{author}{\bibfnamefont{V.}~\bibnamefont{Skumryev}},
  \bibinfo{author}{\bibfnamefont{M.}~\bibnamefont{Varela}},
  \bibinfo{author}{\bibfnamefont{F.}~\bibnamefont{S\'{a}nchez}},
  \bibnamefont{and}
  \bibinfo{author}{\bibfnamefont{J.}~\bibnamefont{Fontcuberta}},
  \bibinfo{journal}{J. Appl. Phys.} \textbf{\bibinfo{volume}{103}},
  \bibinfo{pages}{07E301} (\bibinfo{year}{2008}).

\bibitem[{\citenamefont{L\"{u}ders
  et~al.}(2006{\natexlab{a}})\citenamefont{L\"{u}ders, Herranz, Bibes,
  Bouzehouane, Jacquet, Contour, Fusil, Bobo, Fontcuberta, Barth\'{e}l\'{e}my
  et~al.}}]{Lueders_JAP99_08K301}
\bibinfo{author}{\bibfnamefont{U.}~\bibnamefont{L\"{u}ders}},
  \bibinfo{author}{\bibfnamefont{G.}~\bibnamefont{Herranz}},
  \bibinfo{author}{\bibfnamefont{M.}~\bibnamefont{Bibes}},
  \bibinfo{author}{\bibfnamefont{K.}~\bibnamefont{Bouzehouane}},
  \bibinfo{author}{\bibfnamefont{E.}~\bibnamefont{Jacquet}},
  \bibinfo{author}{\bibfnamefont{J.-P.} \bibnamefont{Contour}},
  \bibinfo{author}{\bibfnamefont{S.}~\bibnamefont{Fusil}},
  \bibinfo{author}{\bibfnamefont{J.-F.} \bibnamefont{Bobo}},
  \bibinfo{author}{\bibfnamefont{J.}~\bibnamefont{Fontcuberta}},
  \bibinfo{author}{\bibfnamefont{A.}~\bibnamefont{Barth\'{e}l\'{e}my}},
  \bibnamefont{et~al.}, \bibinfo{journal}{J. Appl. Phys.}
  \textbf{\bibinfo{volume}{99}}, \bibinfo{pages}{08K301}
  (\bibinfo{year}{2006}{\natexlab{a}}).

\bibitem[{\citenamefont{Chapline and Wang}(2006)}]{Chapline/Wang:2006}
\bibinfo{author}{\bibfnamefont{M.~G.} \bibnamefont{Chapline}} \bibnamefont{and}
  \bibinfo{author}{\bibfnamefont{S.~X.} \bibnamefont{Wang}},
  \bibinfo{journal}{Phys. Rev. B} \textbf{\bibinfo{volume}{74}},
  \bibinfo{pages}{014418} (\bibinfo{year}{2006}).

\bibitem[{\citenamefont{Ramos et~al.}(2008)\citenamefont{Ramos, Santos, Miao,
  Guittet, Moussy, and Moodera}}]{Ramos_PRB78_180402}
\bibinfo{author}{\bibfnamefont{A.~V.} \bibnamefont{Ramos}},
  \bibinfo{author}{\bibfnamefont{T.~S.} \bibnamefont{Santos}},
  \bibinfo{author}{\bibfnamefont{G.~X.} \bibnamefont{Miao}},
  \bibinfo{author}{\bibfnamefont{M.-J.} \bibnamefont{Guittet}},
  \bibinfo{author}{\bibfnamefont{J.-B.} \bibnamefont{Moussy}},
  \bibnamefont{and} \bibinfo{author}{\bibfnamefont{J.~S.}
  \bibnamefont{Moodera}}, \bibinfo{journal}{Phys. Rev. B}
  \textbf{\bibinfo{volume}{78}}, \bibinfo{pages}{180402}
  (\bibinfo{year}{2008}).

\bibitem[{\citenamefont{Suzuki}(2001)}]{Suzuki:2001}
\bibinfo{author}{\bibfnamefont{Y.}~\bibnamefont{Suzuki}},
  \bibinfo{journal}{Annu. Rev. Mater. Res.} \textbf{\bibinfo{volume}{31}},
  \bibinfo{pages}{265} (\bibinfo{year}{2001}).

\bibitem[{\citenamefont{L\"{u}ders et~al.}(2005)\citenamefont{L\"{u}ders,
  Bibes, Bobo, Cantoni, Bertacco, and Fontcuberta}}]{Lueders_PRB71_134419}
\bibinfo{author}{\bibfnamefont{U.}~\bibnamefont{L\"{u}ders}},
  \bibinfo{author}{\bibfnamefont{M.}~\bibnamefont{Bibes}},
  \bibinfo{author}{\bibfnamefont{J.-F.} \bibnamefont{Bobo}},
  \bibinfo{author}{\bibfnamefont{M.}~\bibnamefont{Cantoni}},
  \bibinfo{author}{\bibfnamefont{R.}~\bibnamefont{Bertacco}}, \bibnamefont{and}
  \bibinfo{author}{\bibfnamefont{J.}~\bibnamefont{Fontcuberta}},
  \bibinfo{journal}{Phys. Rev. B} \textbf{\bibinfo{volume}{71}},
  \bibinfo{pages}{134419} (\bibinfo{year}{2005}).

\bibitem[{\citenamefont{Zhou et~al.}(2009)\citenamefont{Zhou, Potzger, Xu,
  Kuepper, Talut, Mark\'{o}, M\"{u}cklich, Helm, Fassbender, Arenholz
  et~al.}}]{Zhou_PRB80_094409}
\bibinfo{author}{\bibfnamefont{S.}~\bibnamefont{Zhou}},
  \bibinfo{author}{\bibfnamefont{K.}~\bibnamefont{Potzger}},
  \bibinfo{author}{\bibfnamefont{Q.}~\bibnamefont{Xu}},
  \bibinfo{author}{\bibfnamefont{K.}~\bibnamefont{Kuepper}},
  \bibinfo{author}{\bibfnamefont{G.}~\bibnamefont{Talut}},
  \bibinfo{author}{\bibfnamefont{D.}~\bibnamefont{Mark\'{o}}},
  \bibinfo{author}{\bibfnamefont{A.}~\bibnamefont{M\"{u}cklich}},
  \bibinfo{author}{\bibfnamefont{M.}~\bibnamefont{Helm}},
  \bibinfo{author}{\bibfnamefont{J.}~\bibnamefont{Fassbender}},
  \bibinfo{author}{\bibfnamefont{E.}~\bibnamefont{Arenholz}},
  \bibnamefont{et~al.}, \bibinfo{journal}{Phys. Rev. B}
  \textbf{\bibinfo{volume}{80}}, \bibinfo{pages}{094409}
  (\bibinfo{year}{2009}).

\bibitem[{\citenamefont{Huang et~al.}(2006)\citenamefont{Huang, Zhu, Zeng, Wei,
  Zhang, and Li}}]{Huang_APL89_265206}
\bibinfo{author}{\bibfnamefont{W.}~\bibnamefont{Huang}},
  \bibinfo{author}{\bibfnamefont{J.}~\bibnamefont{Zhu}},
  \bibinfo{author}{\bibfnamefont{H.~Z.} \bibnamefont{Zeng}},
  \bibinfo{author}{\bibfnamefont{X.~H.} \bibnamefont{Wei}},
  \bibinfo{author}{\bibfnamefont{Y.}~\bibnamefont{Zhang}}, \bibnamefont{and}
  \bibinfo{author}{\bibfnamefont{Y.~R.} \bibnamefont{Li}},
  \bibinfo{journal}{Appl. Phys. Lett.} \textbf{\bibinfo{volume}{89}},
  \bibinfo{pages}{265206} (\bibinfo{year}{2006}).

\bibitem[{\citenamefont{Lisfi et~al.}(2007)\citenamefont{Lisfi, Williams,
  Nguyen, Lodder, Coleman, Corcoran, Johnson, Chang, Kumar, and
  Morgan}}]{Lisfi_PRB76_054405}
\bibinfo{author}{\bibfnamefont{A.}~\bibnamefont{Lisfi}},
  \bibinfo{author}{\bibfnamefont{C.~M.} \bibnamefont{Williams}},
  \bibinfo{author}{\bibfnamefont{L.~T.} \bibnamefont{Nguyen}},
  \bibinfo{author}{\bibfnamefont{J.~C.} \bibnamefont{Lodder}},
  \bibinfo{author}{\bibfnamefont{A.}~\bibnamefont{Coleman}},
  \bibinfo{author}{\bibfnamefont{H.}~\bibnamefont{Corcoran}},
  \bibinfo{author}{\bibfnamefont{A.}~\bibnamefont{Johnson}},
  \bibinfo{author}{\bibfnamefont{P.}~\bibnamefont{Chang}},
  \bibinfo{author}{\bibfnamefont{A.}~\bibnamefont{Kumar}}, \bibnamefont{and}
  \bibinfo{author}{\bibfnamefont{W.}~\bibnamefont{Morgan}},
  \bibinfo{journal}{Phys. Rev. B} \textbf{\bibinfo{volume}{76}},
  \bibinfo{pages}{054405} (\bibinfo{year}{2007}).

\bibitem[{\citenamefont{Gao et~al.}(2009)\citenamefont{Gao, Bao, Birajdar,
  Habisreuther, Mattheis, Schubert, Alexe, and Hesse}}]{Gao_JPhysD42_175006}
\bibinfo{author}{\bibfnamefont{X.~S.} \bibnamefont{Gao}},
  \bibinfo{author}{\bibfnamefont{D.~H.} \bibnamefont{Bao}},
  \bibinfo{author}{\bibfnamefont{B.}~\bibnamefont{Birajdar}},
  \bibinfo{author}{\bibfnamefont{T.}~\bibnamefont{Habisreuther}},
  \bibinfo{author}{\bibfnamefont{R.}~\bibnamefont{Mattheis}},
  \bibinfo{author}{\bibfnamefont{M.~A.} \bibnamefont{Schubert}},
  \bibinfo{author}{\bibfnamefont{M.}~\bibnamefont{Alexe}}, \bibnamefont{and}
  \bibinfo{author}{\bibfnamefont{D.}~\bibnamefont{Hesse}}, \bibinfo{journal}{J.
  Phys. D} \textbf{\bibinfo{volume}{42}}, \bibinfo{pages}{175006}
  (\bibinfo{year}{2009}).

\bibitem[{\citenamefont{L\"{u}ders
  et~al.}(2006{\natexlab{b}})\citenamefont{L\"{u}ders, Barth\'{e}l\'{e}my,
  Bibes, Bouzehouane, Fusil, Jacquet, Contour, Bobo, Fontcuberta, and
  Fert}}]{Lueders_AdvMat18_1733}
\bibinfo{author}{\bibfnamefont{U.}~\bibnamefont{L\"{u}ders}},
  \bibinfo{author}{\bibfnamefont{A.}~\bibnamefont{Barth\'{e}l\'{e}my}},
  \bibinfo{author}{\bibfnamefont{M.}~\bibnamefont{Bibes}},
  \bibinfo{author}{\bibfnamefont{K.}~\bibnamefont{Bouzehouane}},
  \bibinfo{author}{\bibfnamefont{S.}~\bibnamefont{Fusil}},
  \bibinfo{author}{\bibfnamefont{E.}~\bibnamefont{Jacquet}},
  \bibinfo{author}{\bibfnamefont{J.-P.} \bibnamefont{Contour}},
  \bibinfo{author}{\bibfnamefont{J.-F.} \bibnamefont{Bobo}},
  \bibinfo{author}{\bibfnamefont{J.}~\bibnamefont{Fontcuberta}},
  \bibnamefont{and} \bibinfo{author}{\bibfnamefont{A.}~\bibnamefont{Fert}},
  \bibinfo{journal}{Advanced Materials} \textbf{\bibinfo{volume}{18}},
  \bibinfo{pages}{1733} (\bibinfo{year}{2006}{\natexlab{b}}).

\bibitem[{\citenamefont{Rigato et~al.}(2007)\citenamefont{Rigato, Estrad\'{e},
  Arbiol, Peir\'{o}, L\"{u}ders, Mart\'{\i}, S\'{a}nchez, and
  Fontcuberta}}]{Rigato_MatSciEngB144_43}
\bibinfo{author}{\bibfnamefont{F.}~\bibnamefont{Rigato}},
  \bibinfo{author}{\bibfnamefont{S.}~\bibnamefont{Estrad\'{e}}},
  \bibinfo{author}{\bibfnamefont{J.}~\bibnamefont{Arbiol}},
  \bibinfo{author}{\bibfnamefont{F.}~\bibnamefont{Peir\'{o}}},
  \bibinfo{author}{\bibfnamefont{U.}~\bibnamefont{L\"{u}ders}},
  \bibinfo{author}{\bibfnamefont{X.}~\bibnamefont{Mart\'{\i}}},
  \bibinfo{author}{\bibfnamefont{F.}~\bibnamefont{S\'{a}nchez}},
  \bibnamefont{and}
  \bibinfo{author}{\bibfnamefont{J.}~\bibnamefont{Fontcuberta}},
  \bibinfo{journal}{Materials Science and Engineering: B}
  \textbf{\bibinfo{volume}{144}}, \bibinfo{pages}{43} (\bibinfo{year}{2007}).

\bibitem[{\citenamefont{Hohenberg and Kohn}(1964)}]{Hohenberg/Kohn:1964}
\bibinfo{author}{\bibfnamefont{P.}~\bibnamefont{Hohenberg}} \bibnamefont{and}
  \bibinfo{author}{\bibfnamefont{W.}~\bibnamefont{Kohn}},
  \bibinfo{journal}{Phys. Rev.} \textbf{\bibinfo{volume}{136}},
  \bibinfo{pages}{B864} (\bibinfo{year}{1964}).

\bibitem[{\citenamefont{Kohn and Sham}(1965)}]{Kohn/Sham:1965}
\bibinfo{author}{\bibfnamefont{W.}~\bibnamefont{Kohn}} \bibnamefont{and}
  \bibinfo{author}{\bibfnamefont{L.~J.} \bibnamefont{Sham}},
  \bibinfo{journal}{Phys. Rev.} \textbf{\bibinfo{volume}{140}},
  \bibinfo{pages}{A1133} (\bibinfo{year}{1965}).

\bibitem[{\citenamefont{Jones and Gunnarsson}(1989)}]{Jones/Gunnarsson:1989}
\bibinfo{author}{\bibfnamefont{R.~O.} \bibnamefont{Jones}} \bibnamefont{and}
  \bibinfo{author}{\bibfnamefont{O.}~\bibnamefont{Gunnarsson}},
  \bibinfo{journal}{Rev. Mod. Phys.} \textbf{\bibinfo{volume}{61}},
  \bibinfo{pages}{689} (\bibinfo{year}{1989}).

\bibitem[{NiF()}]{NiFe2O4}
\bibinfo{note}{We note that a recent preprint by Ivanov \textit{et al.}
  available at arXiv:1005.2244 reports evidence for a short range $B$ site
  order in NiFe$_2$O$_4$.}

\bibitem[{\citenamefont{N\'{e}el}(1948)}]{Neel}
\bibinfo{author}{\bibfnamefont{L.}~\bibnamefont{N\'{e}el}},
  \bibinfo{journal}{Ann. Phys. (Paris)} \textbf{\bibinfo{volume}{3}},
  \bibinfo{pages}{137} (\bibinfo{year}{1948}).

\bibitem[{\citenamefont{Waldron}(1955)}]{Waldron_PhysRev99_1727}
\bibinfo{author}{\bibfnamefont{R.~D.} \bibnamefont{Waldron}},
  \bibinfo{journal}{Phys. Rev.} \textbf{\bibinfo{volume}{99}},
  \bibinfo{pages}{1727} (\bibinfo{year}{1955}).

\bibitem[{\citenamefont{Jonker}(1959)}]{Jonker_JPhysChemSol9_165}
\bibinfo{author}{\bibfnamefont{G.~H.} \bibnamefont{Jonker}},
  \bibinfo{journal}{J. Phys. Chem. Solids} \textbf{\bibinfo{volume}{9}},
  \bibinfo{pages}{165} (\bibinfo{year}{1959}).

\bibitem[{\citenamefont{P\'{e}nicaud et~al.}(1992)\citenamefont{P\'{e}nicaud,
  Siberchicot, Sommers, and K\"{u}bler}}]{Penicaud_JMagMagMat103_212}
\bibinfo{author}{\bibfnamefont{M.}~\bibnamefont{P\'{e}nicaud}},
  \bibinfo{author}{\bibfnamefont{B.}~\bibnamefont{Siberchicot}},
  \bibinfo{author}{\bibfnamefont{C.~B.} \bibnamefont{Sommers}},
  \bibnamefont{and}
  \bibinfo{author}{\bibfnamefont{J.}~\bibnamefont{K\"{u}bler}},
  \bibinfo{journal}{Journal of Magnetism and Magnetic Materials}
  \textbf{\bibinfo{volume}{103}}, \bibinfo{pages}{212} (\bibinfo{year}{1992}).

\bibitem[{\citenamefont{Antonov et~al.}(2003)\citenamefont{Antonov, Harmon, and
  Yaresko}}]{Antonov_PRB67_024417}
\bibinfo{author}{\bibfnamefont{V.~N.} \bibnamefont{Antonov}},
  \bibinfo{author}{\bibfnamefont{B.~N.} \bibnamefont{Harmon}},
  \bibnamefont{and} \bibinfo{author}{\bibfnamefont{A.~N.}
  \bibnamefont{Yaresko}}, \bibinfo{journal}{Phys. Rev. B}
  \textbf{\bibinfo{volume}{67}}, \bibinfo{pages}{024417}
  (\bibinfo{year}{2003}).

\bibitem[{\citenamefont{Szotek et~al.}(2006)\citenamefont{Szotek, Temmerman,
  K\"{o}dderitzsch, Svane, Petit, and Winter}}]{Szotek_PRB74_174431}
\bibinfo{author}{\bibfnamefont{Z.}~\bibnamefont{Szotek}},
  \bibinfo{author}{\bibfnamefont{W.~M.} \bibnamefont{Temmerman}},
  \bibinfo{author}{\bibfnamefont{D.}~\bibnamefont{K\"{o}dderitzsch}},
  \bibinfo{author}{\bibfnamefont{A.}~\bibnamefont{Svane}},
  \bibinfo{author}{\bibfnamefont{L.}~\bibnamefont{Petit}}, \bibnamefont{and}
  \bibinfo{author}{\bibfnamefont{H.}~\bibnamefont{Winter}},
  \bibinfo{journal}{Phys. Rev. B} \textbf{\bibinfo{volume}{74}},
  \bibinfo{pages}{174431} (\bibinfo{year}{2006}).

\bibitem[{\citenamefont{Zuo et~al.}(2006)\citenamefont{Zuo, Yan, Barbiellini,
  Harris, and Vittoria}}]{Zuo_et_al:2006}
\bibinfo{author}{\bibfnamefont{X.}~\bibnamefont{Zuo}},
  \bibinfo{author}{\bibfnamefont{S.}~\bibnamefont{Yan}},
  \bibinfo{author}{\bibfnamefont{B.}~\bibnamefont{Barbiellini}},
  \bibinfo{author}{\bibfnamefont{V.~G.} \bibnamefont{Harris}},
  \bibnamefont{and} \bibinfo{author}{\bibfnamefont{C.}~\bibnamefont{Vittoria}},
  \bibinfo{journal}{Journal of Magnetism and Magnetic Materials}
  \textbf{\bibinfo{volume}{303}}, \bibinfo{pages}{e432} (\bibinfo{year}{2006}).

\bibitem[{\citenamefont{Perron et~al.}(2007)\citenamefont{Perron, Mellier,
  Domain, Roques, Simoni, Drot, and Catalette}}]{Perron_JPCM19_346219}
\bibinfo{author}{\bibfnamefont{H.}~\bibnamefont{Perron}},
  \bibinfo{author}{\bibfnamefont{T.}~\bibnamefont{Mellier}},
  \bibinfo{author}{\bibfnamefont{C.}~\bibnamefont{Domain}},
  \bibinfo{author}{\bibfnamefont{J.}~\bibnamefont{Roques}},
  \bibinfo{author}{\bibfnamefont{E.}~\bibnamefont{Simoni}},
  \bibinfo{author}{\bibfnamefont{R.}~\bibnamefont{Drot}}, \bibnamefont{and}
  \bibinfo{author}{\bibfnamefont{H.}~\bibnamefont{Catalette}},
  \bibinfo{journal}{J. Phys.: Condens. Matter} \textbf{\bibinfo{volume}{19}},
  \bibinfo{pages}{346219} (\bibinfo{year}{2007}).

\bibitem[{\citenamefont{Jeng and
  Guo}(2002{\natexlab{a}})}]{Jeng_JMagMagMat239_88}
\bibinfo{author}{\bibfnamefont{H.-T.} \bibnamefont{Jeng}} \bibnamefont{and}
  \bibinfo{author}{\bibfnamefont{G.~Y.} \bibnamefont{Guo}},
  \bibinfo{journal}{Journal of Magnetism and Magnetic Materials}
  \textbf{\bibinfo{volume}{239}}, \bibinfo{pages}{88}
  (\bibinfo{year}{2002}{\natexlab{a}}).

\bibitem[{\citenamefont{Jeng and
  Guo}(2002{\natexlab{b}})}]{Jeng_JMagMagMat240_436}
\bibinfo{author}{\bibfnamefont{H.-T.} \bibnamefont{Jeng}} \bibnamefont{and}
  \bibinfo{author}{\bibfnamefont{G.~Y.} \bibnamefont{Guo}},
  \bibinfo{journal}{Journal of Magnetism and Magnetic Materials}
  \textbf{\bibinfo{volume}{240}}, \bibinfo{pages}{436}
  (\bibinfo{year}{2002}{\natexlab{b}}).

\bibitem[{\citenamefont{Kittel}(1949)}]{Kittel_RevModPhys21_541}
\bibinfo{author}{\bibfnamefont{C.}~\bibnamefont{Kittel}},
  \bibinfo{journal}{Rev. Mod. Phys.} \textbf{\bibinfo{volume}{21}},
  \bibinfo{pages}{541} (\bibinfo{year}{1949}).

\bibitem[{\citenamefont{Harrison}(2005)}]{Harrison_QuantumWellsWiresAndDots}
\bibinfo{author}{\bibfnamefont{P.}~\bibnamefont{Harrison}},
  \emph{\bibinfo{title}{Quantum Wells, Wires and Dots: Theoretical and
  Computational Physics of Semiconductor Nanostructures}}
  (\bibinfo{publisher}{Wiley-Interscience}, \bibinfo{year}{2005}),
  \bibinfo{edition}{2nd} ed.

\bibitem[{\citenamefont{Bl\"{o}chl}(1994)}]{Bloechl_PRB50_17953}
\bibinfo{author}{\bibfnamefont{P.~E.} \bibnamefont{Bl\"{o}chl}},
  \bibinfo{journal}{Phys. Rev. B} \textbf{\bibinfo{volume}{50}},
  \bibinfo{pages}{17953} (\bibinfo{year}{1994}).

\bibitem[{\citenamefont{Kresse and Hafner}(1993)}]{Kresse_PRB47_558}
\bibinfo{author}{\bibfnamefont{G.}~\bibnamefont{Kresse}} \bibnamefont{and}
  \bibinfo{author}{\bibfnamefont{J.}~\bibnamefont{Hafner}},
  \bibinfo{journal}{Phys. Rev. B} \textbf{\bibinfo{volume}{47}},
  \bibinfo{pages}{558} (\bibinfo{year}{1993}).

\bibitem[{\citenamefont{Kresse and Hafner}(1994)}]{Kresse_PRB49_14251}
\bibinfo{author}{\bibfnamefont{G.}~\bibnamefont{Kresse}} \bibnamefont{and}
  \bibinfo{author}{\bibfnamefont{J.}~\bibnamefont{Hafner}},
  \bibinfo{journal}{Phys. Rev. B} \textbf{\bibinfo{volume}{49}},
  \bibinfo{pages}{14251} (\bibinfo{year}{1994}).

\bibitem[{\citenamefont{Kresse and
  Furthm\"{u}ller}(1996{\natexlab{a}})}]{Kresse_CompMatSci6_15}
\bibinfo{author}{\bibfnamefont{G.}~\bibnamefont{Kresse}} \bibnamefont{and}
  \bibinfo{author}{\bibfnamefont{J.}~\bibnamefont{Furthm\"{u}ller}},
  \bibinfo{journal}{Comput. Mat. Sci.} \textbf{\bibinfo{volume}{6}},
  \bibinfo{pages}{15} (\bibinfo{year}{1996}{\natexlab{a}}).

\bibitem[{\citenamefont{Kresse and
  Furthm\"{u}ller}(1996{\natexlab{b}})}]{Kresse_PRB54_11169}
\bibinfo{author}{\bibfnamefont{G.}~\bibnamefont{Kresse}} \bibnamefont{and}
  \bibinfo{author}{\bibfnamefont{J.}~\bibnamefont{Furthm\"{u}ller}},
  \bibinfo{journal}{Phys. Rev. B} \textbf{\bibinfo{volume}{54}},
  \bibinfo{pages}{11169} (\bibinfo{year}{1996}{\natexlab{b}}).

\bibitem[{\citenamefont{Anisimov et~al.}(1991)\citenamefont{Anisimov, Zaanen,
  and Andersen}}]{Anisimov/Zaanen/Andersen:1991}
\bibinfo{author}{\bibfnamefont{V.~I.} \bibnamefont{Anisimov}},
  \bibinfo{author}{\bibfnamefont{J.}~\bibnamefont{Zaanen}}, \bibnamefont{and}
  \bibinfo{author}{\bibfnamefont{O.~K.} \bibnamefont{Andersen}},
  \bibinfo{journal}{Phys. Rev. B} \textbf{\bibinfo{volume}{44}},
  \bibinfo{pages}{943} (\bibinfo{year}{1991}).

\bibitem[{\citenamefont{Anisimov et~al.}(1997)\citenamefont{Anisimov,
  Aryasetiawan, and Liechtenstein}}]{Anisimov/Aryatesiawan/Liechtenstein:1997}
\bibinfo{author}{\bibfnamefont{V.~I.} \bibnamefont{Anisimov}},
  \bibinfo{author}{\bibfnamefont{F.}~\bibnamefont{Aryasetiawan}},
  \bibnamefont{and} \bibinfo{author}{\bibfnamefont{A.~I.}
  \bibnamefont{Liechtenstein}}, \bibinfo{journal}{J. Phys.: Condens. Matter}
  \textbf{\bibinfo{volume}{9}}, \bibinfo{pages}{767} (\bibinfo{year}{1997}).

\bibitem[{\citenamefont{Dudarev et~al.}(1998)\citenamefont{Dudarev, Botton,
  Savrasov, Humphreys, and Sutton}}]{Dudarev_PRB57_1505}
\bibinfo{author}{\bibfnamefont{S.~L.} \bibnamefont{Dudarev}},
  \bibinfo{author}{\bibfnamefont{G.~A.} \bibnamefont{Botton}},
  \bibinfo{author}{\bibfnamefont{S.~Y.} \bibnamefont{Savrasov}},
  \bibinfo{author}{\bibfnamefont{C.~J.} \bibnamefont{Humphreys}},
  \bibnamefont{and} \bibinfo{author}{\bibfnamefont{A.~P.}
  \bibnamefont{Sutton}}, \bibinfo{journal}{Phys. Rev. B}
  \textbf{\bibinfo{volume}{57}}, \bibinfo{pages}{1505} (\bibinfo{year}{1998}).

\bibitem[{\citenamefont{Perdew et~al.}(1996)\citenamefont{Perdew, Burke, and
  Ernzerhof}}]{Perdew_PRL77_3865}
\bibinfo{author}{\bibfnamefont{J.~P.} \bibnamefont{Perdew}},
  \bibinfo{author}{\bibfnamefont{K.}~\bibnamefont{Burke}}, \bibnamefont{and}
  \bibinfo{author}{\bibfnamefont{M.}~\bibnamefont{Ernzerhof}},
  \bibinfo{journal}{Phys. Rev. Lett.} \textbf{\bibinfo{volume}{77}},
  \bibinfo{pages}{3865} (\bibinfo{year}{1996}).

\bibitem[{\citenamefont{Li et~al.}(1991)\citenamefont{Li, Fisher, Liu, and
  Nevitt}}]{Li_JMaterSci26_2621}
\bibinfo{author}{\bibfnamefont{Z.}~\bibnamefont{Li}},
  \bibinfo{author}{\bibfnamefont{E.~S.} \bibnamefont{Fisher}},
  \bibinfo{author}{\bibfnamefont{J.~Z.} \bibnamefont{Liu}}, \bibnamefont{and}
  \bibinfo{author}{\bibfnamefont{M.~V.} \bibnamefont{Nevitt}},
  \bibinfo{journal}{J. Materials Science} \textbf{\bibinfo{volume}{26}},
  \bibinfo{pages}{2621} (\bibinfo{year}{1991}).

\bibitem[{\citenamefont{Hill et~al.}(1979)\citenamefont{Hill, Craig, and
  Gibbs}}]{Hill_PhysChemMinerals4_317}
\bibinfo{author}{\bibfnamefont{R.~J.} \bibnamefont{Hill}},
  \bibinfo{author}{\bibfnamefont{J.~R.} \bibnamefont{Craig}}, \bibnamefont{and}
  \bibinfo{author}{\bibfnamefont{G.~V.} \bibnamefont{Gibbs}},
  \bibinfo{journal}{Phys. Chem. Minerals} \textbf{\bibinfo{volume}{4}},
  \bibinfo{pages}{317} (\bibinfo{year}{1979}).

\bibitem[{\citenamefont{Fennie and Rabe}(2006)}]{Fennie_PRL96_205505}
\bibinfo{author}{\bibfnamefont{C.~J.} \bibnamefont{Fennie}} \bibnamefont{and}
  \bibinfo{author}{\bibfnamefont{K.~M.} \bibnamefont{Rabe}},
  \bibinfo{journal}{Phys. Rev. Lett.} \textbf{\bibinfo{volume}{96}},
  \bibinfo{pages}{205505} (\bibinfo{year}{2006}).

\bibitem[{\citenamefont{Neaton et~al.}(2005)\citenamefont{Neaton, Ederer,
  Waghmare, Spaldin, and Rabe}}]{Neaton_PRB71_014113}
\bibinfo{author}{\bibfnamefont{J.~B.} \bibnamefont{Neaton}},
  \bibinfo{author}{\bibfnamefont{C.}~\bibnamefont{Ederer}},
  \bibinfo{author}{\bibfnamefont{U.~V.} \bibnamefont{Waghmare}},
  \bibinfo{author}{\bibfnamefont{N.~A.} \bibnamefont{Spaldin}},
  \bibnamefont{and} \bibinfo{author}{\bibfnamefont{K.~M.} \bibnamefont{Rabe}},
  \bibinfo{journal}{Phys. Rev. B} \textbf{\bibinfo{volume}{71}},
  \bibinfo{pages}{014113} (\bibinfo{year}{2005}).

\bibitem[{\citenamefont{Ederer and Spaldin}(2005)}]{Ederer_CurrOpinion9_128}
\bibinfo{author}{\bibfnamefont{C.}~\bibnamefont{Ederer}} \bibnamefont{and}
  \bibinfo{author}{\bibfnamefont{N.~A.} \bibnamefont{Spaldin}},
  \bibinfo{journal}{Current Opinion in Solid State and Materials Science}
  \textbf{\bibinfo{volume}{9}}, \bibinfo{pages}{128} (\bibinfo{year}{2005}).

\bibitem[{\citenamefont{Terakura et~al.}(1984)\citenamefont{Terakura, Oguchi,
  Williams, and K{\"u}bler}}]{Terakura_PRB30_4734}
\bibinfo{author}{\bibfnamefont{K.}~\bibnamefont{Terakura}},
  \bibinfo{author}{\bibfnamefont{T.}~\bibnamefont{Oguchi}},
  \bibinfo{author}{\bibfnamefont{A.~R.} \bibnamefont{Williams}},
  \bibnamefont{and}
  \bibinfo{author}{\bibfnamefont{J.}~\bibnamefont{K{\"u}bler}},
  \bibinfo{journal}{Phys. Rev. B} \textbf{\bibinfo{volume}{30}},
  \bibinfo{pages}{4734} (\bibinfo{year}{1984}).

\bibitem[{\citenamefont{Lisfi et~al.}(2005)\citenamefont{Lisfi, Williams,
  Johnson, Nguyen, Lodder, Corcoran, Chang, and Morgan}}]{Lisfi_JPCM17_1399}
\bibinfo{author}{\bibfnamefont{A.}~\bibnamefont{Lisfi}},
  \bibinfo{author}{\bibfnamefont{C.~M.} \bibnamefont{Williams}},
  \bibinfo{author}{\bibfnamefont{A.}~\bibnamefont{Johnson}},
  \bibinfo{author}{\bibfnamefont{L.~T.} \bibnamefont{Nguyen}},
  \bibinfo{author}{\bibfnamefont{J.~C.} \bibnamefont{Lodder}},
  \bibinfo{author}{\bibfnamefont{H.}~\bibnamefont{Corcoran}},
  \bibinfo{author}{\bibfnamefont{P.}~\bibnamefont{Chang}}, \bibnamefont{and}
  \bibinfo{author}{\bibfnamefont{W.}~\bibnamefont{Morgan}},
  \bibinfo{journal}{J. Phys.: Condens. Matter} \textbf{\bibinfo{volume}{17}},
  \bibinfo{pages}{1399} (\bibinfo{year}{2005}).

\bibitem[{\citenamefont{Chen et~al.}(1999)\citenamefont{Chen, Snyder,
  Schwichtenberg, Dennis, McCallum, and Jiles}}]{Chen_IEEETransMagn35_3652}
\bibinfo{author}{\bibfnamefont{Y.}~\bibnamefont{Chen}},
  \bibinfo{author}{\bibfnamefont{J.~E.} \bibnamefont{Snyder}},
  \bibinfo{author}{\bibfnamefont{C.~R.} \bibnamefont{Schwichtenberg}},
  \bibinfo{author}{\bibfnamefont{K.~W.} \bibnamefont{Dennis}},
  \bibinfo{author}{\bibfnamefont{R.~W.} \bibnamefont{McCallum}},
  \bibnamefont{and} \bibinfo{author}{\bibfnamefont{D.~C.} \bibnamefont{Jiles}},
  \bibinfo{journal}{IEEE Trans. Magn.} \textbf{\bibinfo{volume}{35}},
  \bibinfo{pages}{3652} (\bibinfo{year}{1999}).

\bibitem[{\citenamefont{Bozorth et~al.}(1955)\citenamefont{Bozorth, Tilden, and
  Williams}}]{Bozorth_PhysRev99_1788}
\bibinfo{author}{\bibfnamefont{R.~M.} \bibnamefont{Bozorth}},
  \bibinfo{author}{\bibfnamefont{E.~F.} \bibnamefont{Tilden}},
  \bibnamefont{and} \bibinfo{author}{\bibfnamefont{A.~J.}
  \bibnamefont{Williams}}, \bibinfo{journal}{Phys. Rev.}
  \textbf{\bibinfo{volume}{99}}, \bibinfo{pages}{1799} (\bibinfo{year}{1955}).

\bibitem[{\citenamefont{Smith and Jones}(1966)}]{Smith_JAP37_1001}
\bibinfo{author}{\bibfnamefont{A.~B.} \bibnamefont{Smith}} \bibnamefont{and}
  \bibinfo{author}{\bibfnamefont{R.~V.} \bibnamefont{Jones}},
  \bibinfo{journal}{J. Appl. Phys.} \textbf{\bibinfo{volume}{37}},
  \bibinfo{pages}{1001} (\bibinfo{year}{1966}).

\end{thebibliography}

\end{document}